\documentclass[%
 aps, prl,
 amsmath,amssymb,
 reprint,%
superscriptaddress
]{revtex4-2}

\usepackage{graphicx}% Include figure files
\usepackage{bm}% bold math
\usepackage{fixme}
\usepackage{hyperref}
\usepackage[utf8]{inputenc}
\usepackage[T1]{fontenc}
\usepackage{mathptmx}
\usepackage{lipsum}
\usepackage{amsmath}
\usepackage{physics}
\usepackage{xparse}
\usepackage{bbm}
\usepackage{xcolor}
\usepackage{url}

\graphicspath{{Pictures/}}

\newcommand{\mytitile}{Photon transport in a Bose-Hubbard chain of superconducting artificial atoms}

\begin{document}
	\preprint{AIP/123-QED}
	
	\title[\mytitile]{\mytitile\\~}
	\author{G.P. Fedorov}
	\email{gleb.fedorov@phystech.edu}
	
	\affiliation{ 
		Moscow Institute of Physics and Technology, Dolgoprundiy, Russia
	}
	\affiliation{ 
		Russian Quantum Center, National University of Science and Technology MISIS, 119049 Moscow, Russia
	}%
	\affiliation{National University of Science and Technology MISIS, 119049 Moscow, Russia}

	\author{S. V. Remizov}
	\affiliation{Dukhov Automatics Research Institute, (VNIIA), 127055 Moscow, Russia}
	\affiliation{Kotel'nikov Institute of Radio Engineering and Electronics, Russian Academy of Sciences, 125009 Moscow, Russia}
	\author{D. S. Shapiro}
	\affiliation{Dukhov Automatics Research Institute, (VNIIA), 127055 Moscow, Russia}

	\affiliation{Kotel'nikov Institute of Radio Engineering and Electronics, Russian Academy of Sciences, 125009 Moscow, Russia}
	\author{W. V. Pogosov}
	\affiliation{Dukhov Automatics Research Institute, (VNIIA), 127055 Moscow, Russia}
	
	\affiliation{Institute for Theoretical and Applied Electrodynamics, Russian Academy of Sciences, 125412 Moscow, Russia}

	\author{E. Egorova}

\affiliation{ 
	Moscow Institute of Physics and Technology, Dolgoprundiy, Russia
}
\affiliation{ 
Russian Quantum Center, National University of Science and Technology MISIS, 119049 Moscow, Russia
}%
	\affiliation{National University of Science and Technology MISIS, 119049 Moscow, Russia}

	\author{I. Tsitsilin}

\affiliation{ 
	Moscow Institute of Physics and Technology, Dolgoprundiy, Russia
}
\affiliation{ 
Russian Quantum Center, National University of Science and Technology MISIS, 119049 Moscow, Russia
}%
	\affiliation{National University of Science and Technology MISIS, 119049 Moscow, Russia}
	
	\author{M. Andronik}
	\affiliation{FMN Laboratory, Bauman Moscow State Technical University, 105005 Moscow, Russia}
	
	\author{A. A. Dobronosova}
	
	\affiliation{FMN Laboratory, Bauman Moscow State Technical University, 105005 Moscow, Russia}
	\affiliation{Dukhov Automatics Research Institute, (VNIIA), 127055 Moscow, Russia}
	
	\author{I. A. Rodionov}
	
	\affiliation{FMN Laboratory, Bauman Moscow State Technical University, 105005 Moscow, Russia}
	\affiliation{Dukhov Automatics Research Institute, (VNIIA), 127055 Moscow, Russia}
	
	\author{O.V. Astafiev}
	\affiliation{Skolkovo Institute of Science 
		and Technology, Moscow, Russia}
	\affiliation{ 
		Moscow Institute of Physics and Technology, 
		Dolgoprundiy, Russia
	}
	\affiliation{Physics Department, Royal 
	Holloway, University of London, Egham, Surrey 
	TW20 0EX, United Kingdom}
\affiliation{National Physical Laboratory, Teddington, TW11 0LW, United Kingdom}

	\author{A.V. Ustinov}
	\affiliation{ 
		Russian Quantum Center, National University of Science and Technology MISIS, 119049 Moscow, Russia
	}
	\affiliation{Physics Institute and Institute for Quantum Materials and Technologies, Karlsruhe Institute of Technology, 76131 Karlsruhe, Germany}
	\affiliation{National University of Science and Technology MISIS, 119049 Moscow, Russia}
	
	\date{\today}% It is always \today, today,

	\begin{abstract}
We demonstrate non-equilibrium steady-state photon transport through a chain of five coupled artificial atoms simulating the driven-dissipative Bose-Hubbard model. Using transmission spectroscopy, we show that the system retains many-particle coherence despite being coupled strongly to two open spaces. We show that system energy bands may be visualized with high contrast using cross-Kerr interaction. For vanishing disorder, we observe the transition of the system from the linear to the nonlinear regime of photon blockade in excellent agreement with the input-output theory. Finally, we show how controllable disorder introduced to the system suppresses this non-local photon transmission. We argue that proposed architecture may be applied to analog simulation of many-body Floquet dynamics with even larger arrays of artificial atoms paving an alternative way to demonstration of quantum supremacy.
\end{abstract}
	
\maketitle

There has been increased effort over recent years in on-chip simulation of various solid state and quantum optical models using superconducting circuits \cite{kjaergaard2019superconducting}. The Bose-Hubbard (B-H) model is now particularly well-covered as it can be straightforwardly mapped onto arrays of coupled transmon qubits \cite{orell2019probing,yanay2020two}. The pioneering work \cite{hacohen2015cooling} had demonstrated this for a three-site linear lattice, and subsequent experiments were focused on simulating dynamics with engineered dissipation \cite{ma2019dissipatively}, investigating the many-body localization phase transitions \cite{roushan2017spectroscopic,chiaro2019growth}, and correlated quantum walks \cite{Yan2019, Ye2019}. As numerous theoretical studies propose a new research direction involving controllable light-matter interaction and Floquet engineering to study periodically-driven Hamiltonians and their non-equilibrium dynamics \cite{Goldman2014, eisert2015quantum, Zippilli2015, kyriienko2018floquet, franca2020simulating}, it is tempting to use transmon chains  to simulate the driven-dissipative Bose-Hubbard model. The subject is particularly interesting since a recent study has shown that driven systems may open new ways to demonstrate quantum supremacy \cite{tangpanitanon2019quantum}.

In this Letter, we present a proof-of-principle device which models non-equilibrium steady-state boson transport through a Bose-Hubbard chain using a linear array of five transmons strongly coupled to waveguides at its edges. While dominating over other loss channels, this strong coupling is still negligible compared to the interaction between the transmons and thus does not destroy the many-body coherence of the system. This means that such architecture with an increased number of transmons could be suitable for supremacy-scale Floquet quantum simulations. Moreover, this device complements previous theoretical research on the transmission spectroscopy of quantum metamaterials \cite{Zagoskin2016, viehmann2013observing, Greenberg2015, Fistul2019, Biella2015,roberts2020driven, collodo2019observation,tiwari2020interplay} with direct experimental data. We also expect that similar systems may be used to test the accuracy of methods of contraction of the Hilbert space such as the matrix product states or tensor networks in general \cite{Biella2015, orell2019probing, di2019efficient}.

\begin{figure}
	\centering
	\includegraphics[width=1\linewidth]{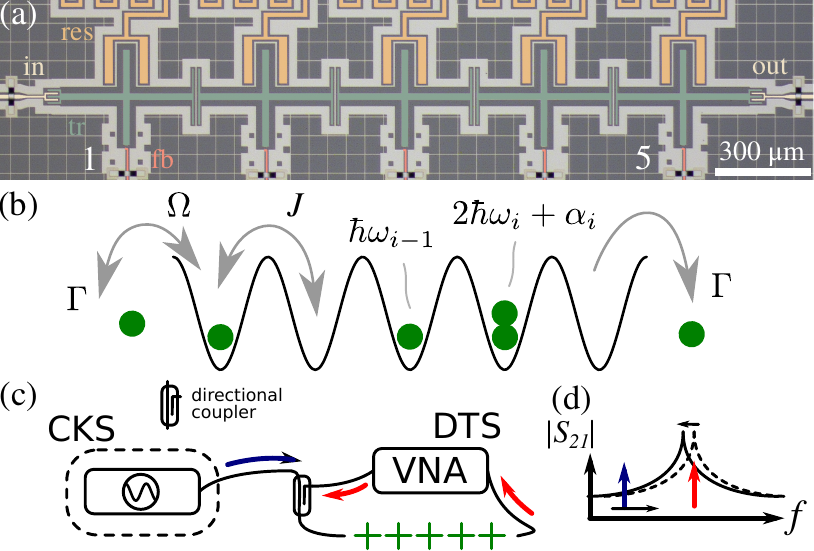}
	\caption{\textbf{(a)} Optical image of the device (false-colored). Input and output waveguides (beige) are strongly coupled to the edge transmons (green). All transmons can be dispersively read out via auxiliary resonators (orange) and tuned via flux bias lines (red). \textbf{(b)} Model of the device -- a B-H lattice with five sites. Bosons are inserted from the left by the drive of strength $\Omega$, and leak predominantly from the sides at rate $\Gamma$. The energy of an i\textsuperscript{{th}} localized boson is $\hbar \omega_i$, adding another boson to the same site costs $\hbar \alpha_i < 0$. Bosons can tunnel between sites at rate $J$. \textbf{(c)} Qualitative measurement schematic: the direct transmission spectroscopy (DTS) is done using the vector network analyzer which measures the complex transmission $ S_{21} $. The cross-Kerr spectroscopy (CKS) requires an additional microwave source connected through a directional coupler. \textbf{(d)} CKS is done by sweeping the source frequency (blue arrow) while monitoring the transmission at a certain resonance peak via the VNA (red arrow).}
	\label{fig:scheme}
\end{figure}

The layout of the chip is shown in \autoref{fig:scheme}~(a). We use five capacitively coupled Xmon qubits tunable via individual flux lines. Strong coupling to the open spaces is attained via large interdigitated capacitors at the input and output waveguides and allows to measure the microwave transmission through the system. In \autoref{fig:scheme}~(b), we illustrate the physical model simulated by the device; the corresponding Hamiltonian including the coherent drive is
\begin{equation}
\begin{aligned}
\hat H/\hbar &= \sum_{i=1}^5\left[ (\omega_i - \omega_d) \hat b^\dag_i \hat b_i + \frac{1}{2} \alpha_i \hat b_i^\dag \hat b_i (\hat b^\dag_i \hat b_i - 1)\right]\\
&+J\sum_{i=1}^4 \left[\hat b^\dag_{i+1} \hat b_i + \hat b_{i+1} \hat b_i^\dag\right]+\frac{\Omega}{2}(\hat b_1^\dag + \hat b_1),
\end{aligned}\label{eq:bose-hubbard}
\end{equation} 
where $\hat b_i$, $\hbar \omega_i$ and $\hbar\alpha_i$ are, respectively, site lowering operator, single-boson energy, on-site interaction for the $i$\textsuperscript{th} site; $J$ is the site-site tunneling rate, and $\omega_d$ is the drive frequency \cite{egorova2020analog, PhysRevA.102.013707, yanay2020two}. The dissipation essential to the dynamics is included in the corresponding Liouville equation using Lindbladian superoperators $\mathcal D[\hat{O}^{(i)}_\alpha] = \hat{O}^{(i)}_\alpha \hat \rho \hat{O}^{(i)\dag}_\alpha - \frac{1}{2}\{\hat{O}^{(i)\dag}_\alpha \hat{O}^{(i)}_\alpha, \hat \rho\},$ where $\hat{O}^{(i)}_\gamma = \sqrt{\gamma_i} \hat b_i$ is the relaxation and $\hat{O}^{(i)}_\phi = \sqrt{\gamma^{(i)}_\phi} \hat b_i^\dag \hat b_i$ is the pure dephasing. Strong coupling to the edge lines implies $\gamma_1 \approx \gamma_5 \approx \Gamma$ to be the dominating source of decoherence. If $\omega_i = \omega,\ \alpha_i = \alpha$, the standard B-H Hamiltonian is restored.

In \autoref{fig:scheme}~(c) we show schematically the experimental setup. We measure the transmission $S_{21}$ through the chain using a vector network analyzer (direct transmission spectroscopy, DTS) and optionally use an additional microwave source to perform the cross-Kerr spectroscopy (CKS) of the system; in both cases, with the continuous microwave excitation we study the steady-state properties of the device. To obtain theoretical predictions for the $S_{21}$ in DTS, one can use the input-output formalism \cite{yurke1984quantum,gardiner1985input}. Since we irradiate the system coherently, we assume that the input field mode amplitude is related to the coherent drive strength $\Omega$ in the driving operator $\hbar \Omega \hat b_1 \cos \omega t$ via $\sqrt{\gamma_1} \langle  \hat b_{in}^\dag \rangle \approx i \Omega/2$, which follows from the quantum Langevin equations \cite{mirhosseini2019cavity}. Similarly, the output field operator $\hat b_{out}^\dag \approx \sqrt{\gamma_5} \hat b_5^\dag$. From this, we obtain $
	S_{21} = \langle \hat b_{out}^\dag\rangle / \langle \hat b_{in}^\dag \rangle = 2\Gamma\cdot \Tr[\hat \rho_{ss} \hat b_5^\dag]/{i\Omega} ,
$
where $\hat \rho_{ss}$ is the steadystate density matrix. Physically, this expression means that the signal transmission is possible if the rightmost transmon becomes non-locally excited while the leftmost is subject to radiation. Indeed, from the linearized Langevin equations \cite{astafiev2010resonance} ($\Omega \ll \Gamma$) follows that  and at the degeneracy point ($\omega_i = \omega$) five transmission peaks detuned $0,\, \pm J,\, \pm \sqrt{3} J$ from $\omega$ should appear due to the interaction, corresponding to the classical normal mode frequencies. The widths of the central, next-to-central and edge peaks are $2\Gamma/3,\ \Gamma/2$ and $\Gamma/6$, respectively, and add up to $2\Gamma$ (see Supplemental Material).  In the quantum-mechanical limit, these resonances should remain in the spectrum due to the correspondence principle; however, new lines caused by purely quantum-mechanical processes are expected to appear in the nonlinear regime.

\begin{figure}[b]
	\centering
	\includegraphics[width=1\linewidth]{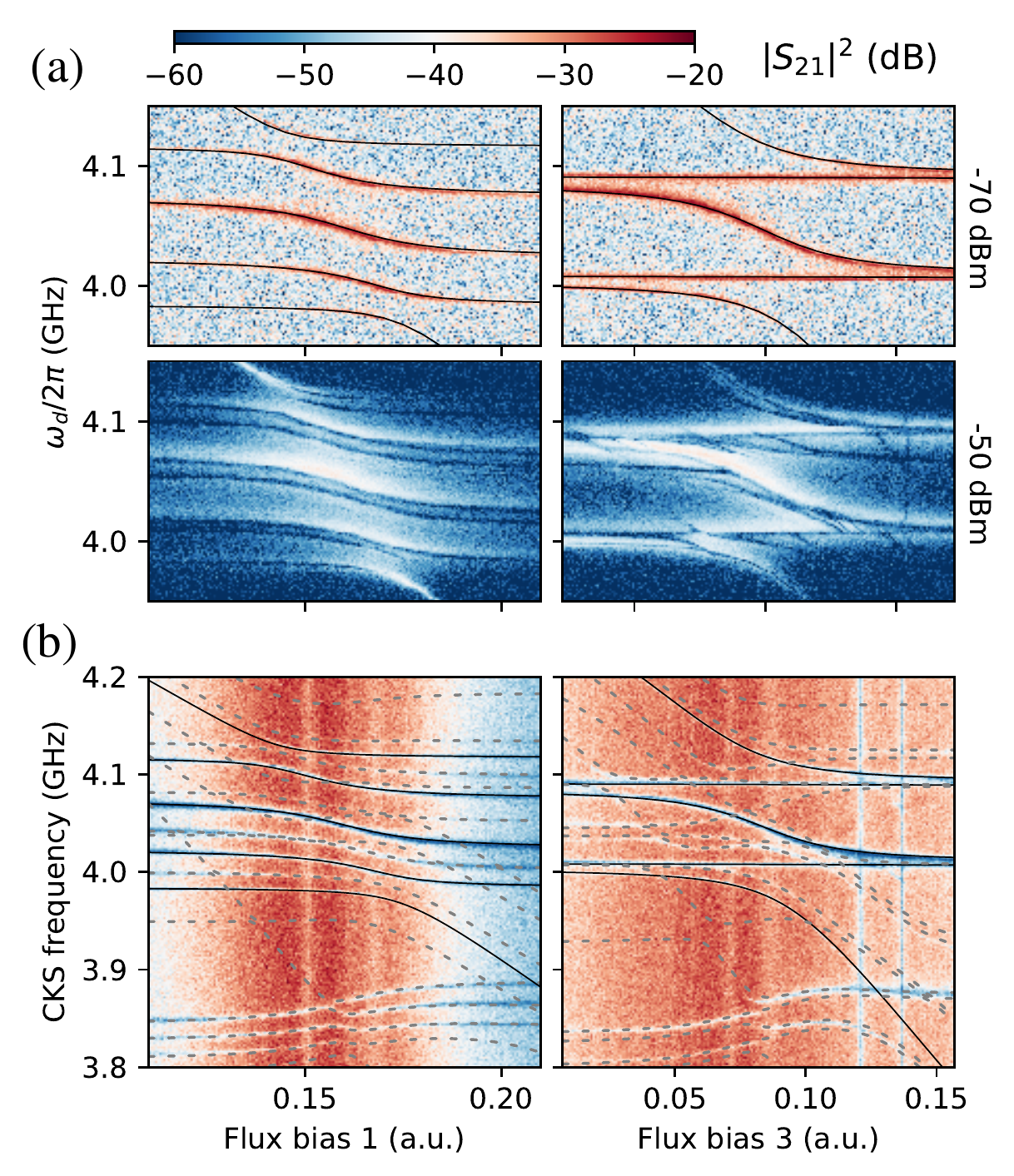}
	
	\caption{\textbf{(a)} DTS of the chain. $S_{21}$ includes the attenuation and amplification in the measurement chain, VNA output power is shown on the right. In the top row, black lines are the fit of the lowest five transitions of \autoref{eq:bose-hubbard} for $J/2\pi = 41$ MHz, $\omega_i/2\pi \approx 4.05$ GHz. \textbf{(b)} CKS via the the third mode showing the emergent band structure of the system. Dashed lines are fits to the purely quantum transitions.}
	\label{fig:transmission}
\end{figure}

\begin{figure*}
	\centering
	\includegraphics[width=\linewidth]{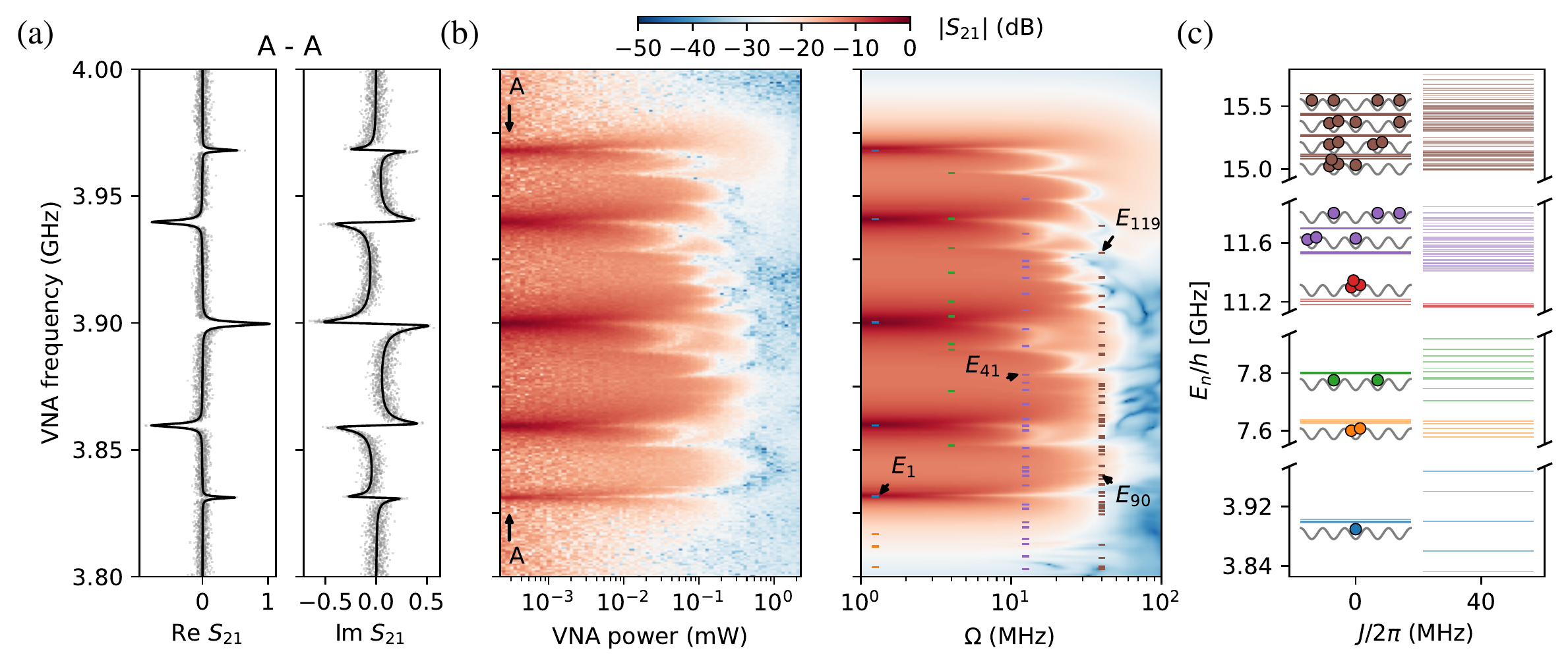}

	\includegraphics[width=.33\linewidth]{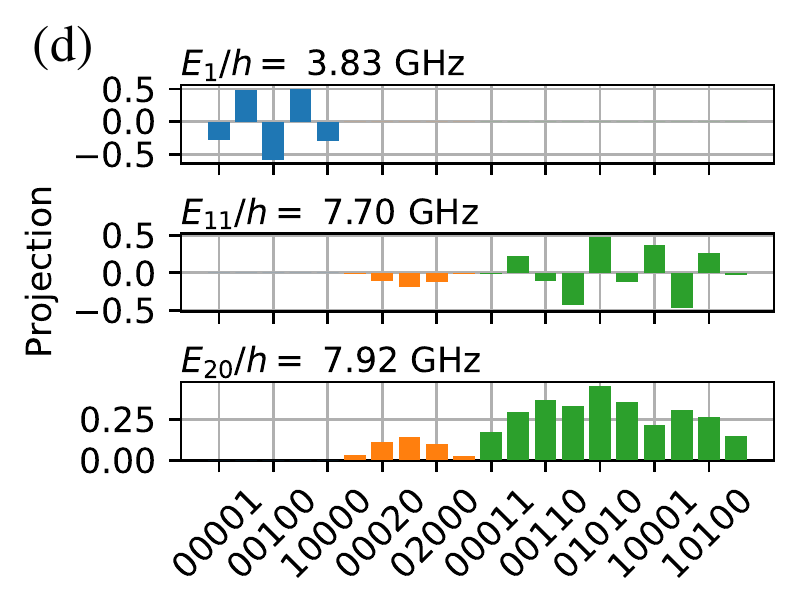}
	\includegraphics[width=.33\linewidth]{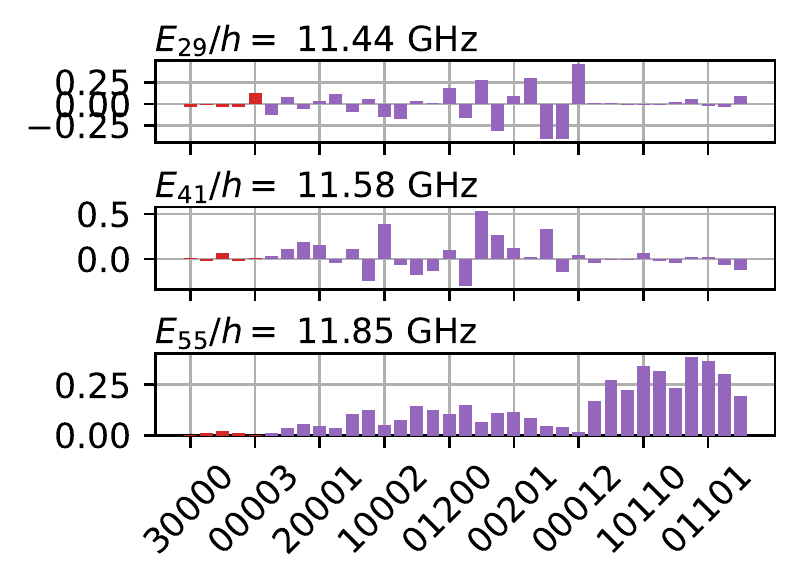}
	\includegraphics[width=.33\linewidth]{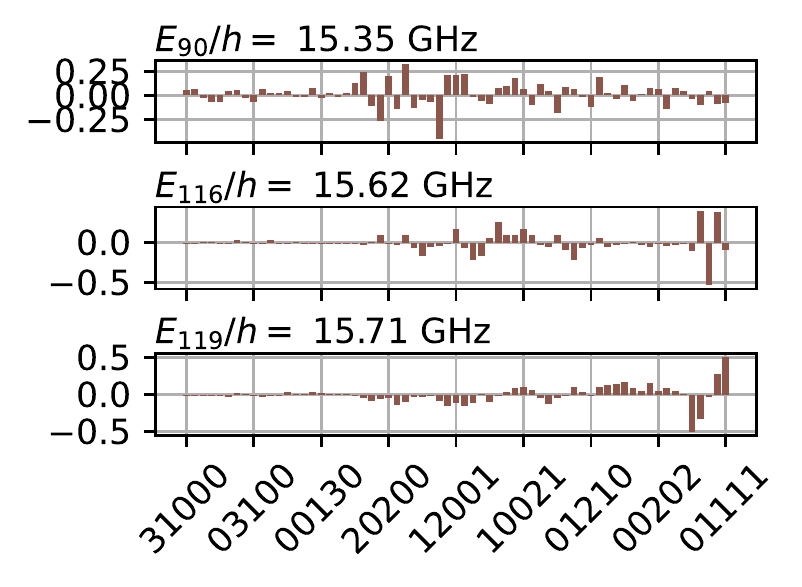}
	\caption{\textbf{(a)} The analytical solution for the $S_{21}$ in the linear regime (smooth curves) fitted to the low power data (clouds), normalized. \textbf{(b)} Experimental and simulated $|S_{21}|$ for various driving powers. Driving power is calibrated to match the Rabi frequency $\Omega$ of the simulation.  Dashes show the corresponding multiphoton transition frequencies calculated using the eigenlevels from (c). \textbf{(c)} The energy level structure of the model with and without interaction using the parameters extracted from the fits; three excitations per site are included, and up to four excitations total; $E_n$ is the energy of the $n$-th eigenstate, $E_0 = 0$. \textbf{(d)} Relevant many-body eigenstates projected onto the unperturbed basis (colors as in (c)). The inherent randomness in the decompositions of the high-energy states conditions the random structure of the energy levels.}
	\label{fig:cq_transition}
\end{figure*}

The results of the DTS are shown in \autoref{fig:transmission}~(a). To show the structure of the eigenmodes and to extract $\omega_i$ and $J$, we bias one of the transmons across the degeneracy point while keeping the others at around 4.05 GHz. In the left column, the first transmon is swept, and in the right, the middle one. The first transmon interacts with all collective states, and the third only with the odd ones; this behavior is expected from the Hamiltonian. We find the tunnelling rate $J/2\pi$ to be around 41 MHz from the numerical fit. When the incident power is increased, the resonances are subject to photon blockade \cite{birnbaum2005photon} and behave similarly to what is observed for single superconducting qubits \cite{astafiev2010resonance}. In the bottom row of \autoref{fig:transmission}~(a), we notice spectral manifestations of the many-body states of the system which do not have any classical analogy and cannot be observed in a non-composite quantum system. We thus call this power-dependent behavior shown in \autoref{fig:transmission}~(a) the classical-quantum transition.

The remaining parameters $\alpha_i$ of \autoref{eq:bose-hubbard} can be extracted via the CKS. In \autoref{fig:transmission}~(b) we have done it using the same two configurations of the transmon frequencies as in \autoref{fig:transmission}~(a) and performed another numerical fit (solid and dashed lines). The readout tone was aimed at the third mode, so the observed spectral frequencies should be corrected by adding its frequency for each bias voltage. The dashed lines show the emergent bands of the two-photon subspace: the many-body states with two excitations at different sites are near 4.05 GHz and ``doublons'' \cite{gorlach2018simulation} are located around 3.85 GHz. The B-H eigenstates with doubly-populated sites have lower energy due to the attractive interactions; the disorder in the extracted values of $\alpha_i/2\pi$ of approximately [-188, -178, -178, -178, -188] MHz is around 5\% and is caused by the uncompensated capacitance to the transmission lines. 
%\section{Classical to quantum transition}
%\begin{figure*}[t]
%	\centering
%	\caption{}
%	\label{fig:eigenstates}
%\end{figure*}

To further study the energy structure and the non-equilibrium dynamics of the system during the classical-quantum transition, we use a direct transmission experiment with fixed degenerate configuration of the transmon frequencies $\omega_i/2\pi = 3.9$ GHz. Using a fitting procedure similar to \autoref{fig:transmission} we find $\omega_i/2\pi$ to [3.898, 3.898, 3.9, 3.901, 3.901] MHz where differences from the target value come from the flux cross-talk. In the linear regime, we estimate the coupling to the transmission lines and internal dissipation from the fit of the complex transmission coefficient predicted by the linear model which is shown in \autoref{fig:cq_transition}~(a). Using these data, we also estimate the transmission amplitude through the attenuation and amplification chain and find that the third mode has nearly unity transmission. This is expected, as from \autoref{fig:transmission}~(a) it is only coupled to a single ``bulk'' transmon, and thus has the least internal dissipation. Since in the linear model it is impossible to discern pure dephasing and internal dissipation, the relaxation rates from the fit are larger than true values: we estimate $\gamma_i$ to be [16, 6, 0.1,  3, 16] $\mu\text{s}^{-1}$; the rates $\gamma_1$ and $\gamma_5$ are in good agreement with the value calculated from the simulated edge capacitances of 8 fF and justify the assumption of dominating $\Gamma$.

\autoref{fig:cq_transition}~(b) shows how the transmission spectrum changes throughout the transition. The normal mode peaks gradually saturate due to the photon blockade and multiple new dips appear caused by the reflective multiphoton transitions to the many-body eigenstates \cite{Biella2015,PhysRevA.102.013707,roberts2020driven}. The experimental data agrees very well with the numerical steady-state simulation in \textit{qutip} \cite{qutip1, qutip2} of the five-site Bose-Hubbard model with the parameters extracted earlier and three bosons per site at max; full simulation of 253$\times$253 density matrix takes approximately a week on a 138 core cluster for the shown 300$\times$300 heatmap. The selection rules of the system do not allow all possible multiphoton lines, but one can clearly discern an increase of the density of states and their randomness with increasing band number. It can be connected to the classical chaotisity of the system if the distribution of the level spacings corresponds to the Gaussian orthogonal ensemble \cite{bohigas1984characterization,zimmermann1986manifestation, livan2018introduction}. The frequencies of transitions up to four-photon are shown with colored dashes and can be identified with the energy bands shown in \autoref{fig:cq_transition}~(c) calculated using the fitted parameters of the model \autoref{eq:bose-hubbard}. The statistics of the calculated nearest-neighbor level spacings resembles the Wigner-Dyson distribution for the experimentally determined parameters, but is rather Poissonian for the ideal parameters which probably means that the system size is too low to obtain the correct histogram. To show how delocalized are the eigenstates $ \ket{n} $ reachable in our device ($n$ is counted excluding the states with more than three excitation per site), in \autoref{fig:cq_transition}~(d) we project several of them onto the non-interacting basis. State $\ket{1}$ has the structure identical to a classical symmetric low-frequency eigenmode. States $\ket{11}$ and $\ket{20}$ are at the edges of the orange band of the two-photon subspace and are much larger superpositions. One can note the unveiling randomness in the decomposition coefficients, which becomes more and more pronounced for higher energies: no symmetries or even any kind of structure can be found in the higher eigenstates except for the hole-like four-excitation subspace dual to the single-photon one (see $\ket{116}, \ket{119}$). 

\begin{figure}
	\includegraphics[width=1\linewidth]{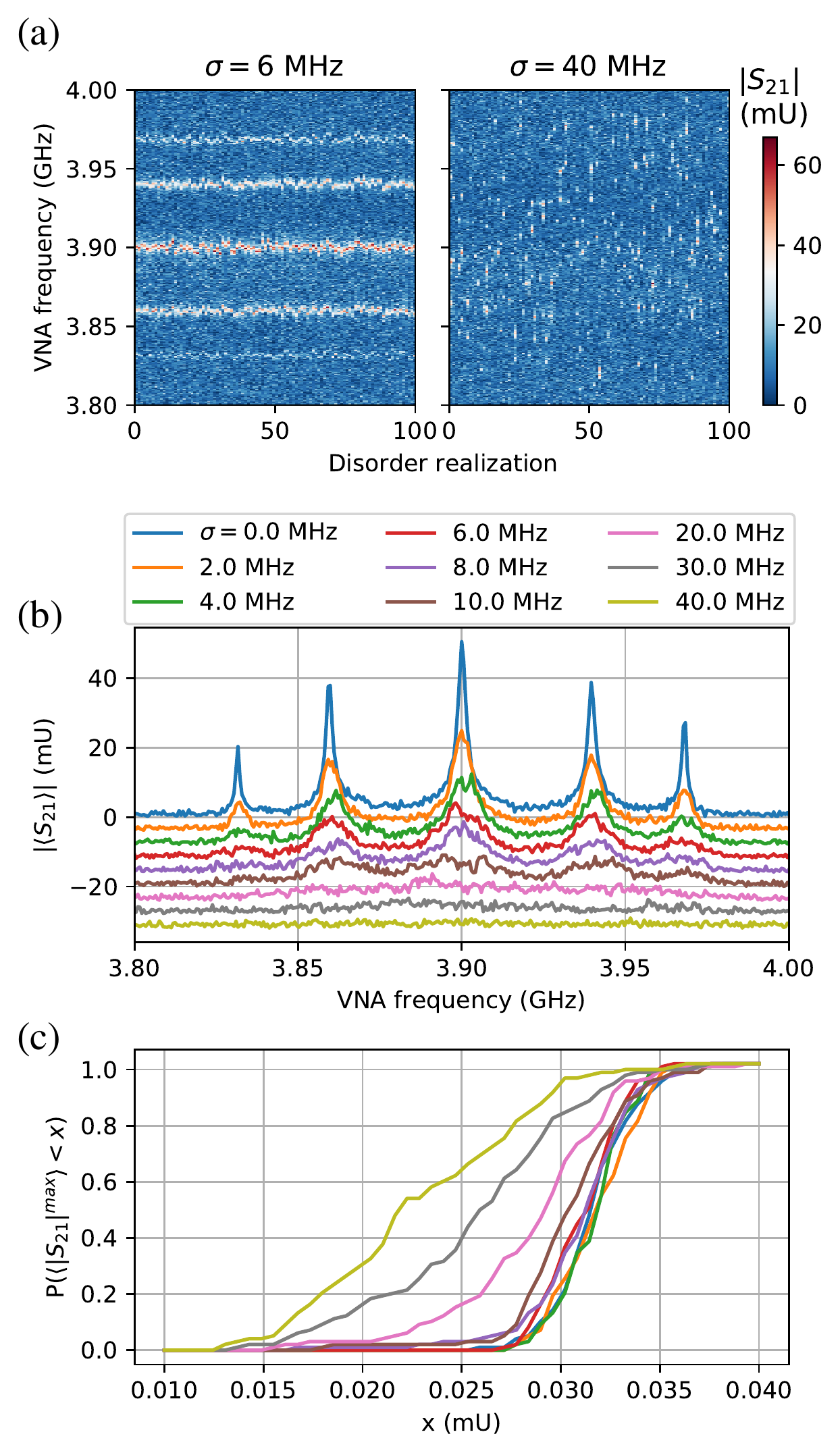}
	\caption{\textbf{(a)} Raw transmission data for $\sigma = 6,\, 40$ MHz and 100 realizations of disorder; absolute value of the transmission is shown with color. \textbf{(b)} Localization and disappearance of transmission with increasing disorder. Each curve shows the absolute value of the averaged transmission over the Gaussian disorder realizations with a certain standard deviation $\sigma$ shown in the legend. Each next curve is offset downwards for better visibility. \textbf{(c)} Probability distributions of the brightest peak prominence $\langle |S_{21}|^{max}\rangle$ (see text) observed in the disorder realizations for the $\sigma$ values from (b).}
	\label{fig:mbl}
\end{figure}

It is known that the Poissoinian statistics is usually a property of the disordered Bose-Hubbard model exhibiting localization \cite{roushan2017spectroscopic, Yan2019,Ye2019}. To check how the localization changes the transport properties, we introduce controllable disorder into the transmon frequencies near the degeneracy point in an experiment similar to what was done before numerically \cite{orell2019probing}: a certain common frequency variance $\sigma$ is chosen, then the random frequency $\omega + \delta \omega$ is assigned to each transmon where $\delta\omega \in \mathbb N(0,2\pi\sigma)$ and $\omega/2\pi = 3.9$ GHz. Then the transmission is recorded, and the full process is repeated 100 times. 
In \autoref{fig:mbl}~(a) we show two examples of the raw transmission data for  $\sigma = 6,\, 40$ MHz. As one can see, for the smaller standard deviation of the target frequencies, the eigenmodes stay relatively unchanged while for the larger the initial structure is completely lost. The averaged curves for several values of $\sigma$ are shown in \autoref{fig:mbl}~(b). One can see that when the noise in the transmon frequencies reaches the coupling strength $J/2\pi$, the averaged transmission vanishes. This means that the localization is revealed in the transport properties when the excitation of the first qubit on average does not reach the last qubit. This fact reminds of the superconductor-insulator transition \cite{bruder1993superconductor} to describe which was the initial purpose of the Bose-Hubbard model. As the transmission vanishes only on average and some peaks occasionally remain even for the largest $\sigma$, in \autoref{fig:mbl}~(c) we also study the distribution of the brightest peak prominences (taken as the mean of 10 points around the maximum) seen over the disorder realizations and show that it also changes qualitatively when $\sigma \approx J/2\pi$.  

%As we only study the lowest excited modes of the system in the linear regime, one can argue that this effect may be equally well described by the classical coupled mode theory. Similarly, the thermalization of a system of coupled linear oscillators would not occur when a single normal \textit{mode} is excited \cite{deutsch2018eigenstate}, but when a single \textit{oscillator} is, the behavior of such system should be identical to the one tested for chains of coupled transmons before \cite{Yan2019, ma2019dissipatively} for low disorder configurations and single-photon states; the observed values would then be not the populations of the single-transmon Fock-states but the amplitudes of the oscillations at each site. However, this is only a shallow analogy showing the continuity between classical and quantum physics \cite{park2012classical}.

In conclusion, we have shown how quantum photon transport occurs through a Bose-Hubbard chain simulated by transmon artificial atoms. We have demonstrated that the behavior of the single-photon subspace of the system does not deviate from the classical normal mode theory which is expected from the correspondence principle \cite{park2012classical}. However, an increase of the incident photon flux beyond the dissipation rate reveals the quantum nature of the system through the photon blockade and multiphoton transitions to composite many-body states. The classical theory then fails, and one can only resort to numerical solution of the master equation to find the non-equilibrium steady state, which shows excellent agreement with the data. Finally, we have shown how controllable disorder affects the photon transport: we find that the transmission averaged over disorder realizations ceases when the standard deviation of the transmon frequencies reaches the interaction strength.

We gratefully acknowledge valuable discussions with I.S. Besedin, S. Flach and A. Poddubny. We thank D. Yakovlev and A. Sokolova for their help in preliminary experiments. The investigation was conducted with the support of Russian Science Foundation, Grants No. 16-12-00070 (measurement and data analysis) and 16-12-00095 (numerical modeling). Devices were fabricated at the BMSTU Nanofabrication Facility (Functional Micro/Nanosystems, FMNS REC, ID 74300). We also acknowledge support from the Ministry of Education and Science of the Russian Federation in the framework of the Increased Competitiveness Program of the National University of Science and Technology MISIS (Contract No. K2-2020-022). 

\bibliographystyle{apsrev4-1}
\bibliography{abbreviated}% Produces the bibliography via BibTeX.	

\begin{thebibliography}{39}%
\makeatletter
\providecommand \@ifxundefined [1]{%
 \@ifx{#1\undefined}
}%
\providecommand \@ifnum [1]{%
 \ifnum #1\expandafter \@firstoftwo
 \else \expandafter \@secondoftwo
 \fi
}%
\providecommand \@ifx [1]{%
 \ifx #1\expandafter \@firstoftwo
 \else \expandafter \@secondoftwo
 \fi
}%
\providecommand \natexlab [1]{#1}%
\providecommand \enquote  [1]{``#1''}%
\providecommand \bibnamefont  [1]{#1}%
\providecommand \bibfnamefont [1]{#1}%
\providecommand \citenamefont [1]{#1}%
\providecommand \href@noop [0]{\@secondoftwo}%
\providecommand \href [0]{\begingroup \@sanitize@url \@href}%
\providecommand \@href[1]{\@@startlink{#1}\@@href}%
\providecommand \@@href[1]{\endgroup#1\@@endlink}%
\providecommand \@sanitize@url [0]{\catcode `\\12\catcode `\$12\catcode
  `\&12\catcode `\#12\catcode `\^12\catcode `\_12\catcode `\%12\relax}%
\providecommand \@@startlink[1]{}%
\providecommand \@@endlink[0]{}%
\providecommand \url  [0]{\begingroup\@sanitize@url \@url }%
\providecommand \@url [1]{\endgroup\@href {#1}{\urlprefix }}%
\providecommand \urlprefix  [0]{URL }%
\providecommand \Eprint [0]{\href }%
\providecommand \doibase [0]{http://dx.doi.org/}%
\providecommand \selectlanguage [0]{\@gobble}%
\providecommand \bibinfo  [0]{\@secondoftwo}%
\providecommand \bibfield  [0]{\@secondoftwo}%
\providecommand \translation [1]{[#1]}%
\providecommand \BibitemOpen [0]{}%
\providecommand \bibitemStop [0]{}%
\providecommand \bibitemNoStop [0]{.\EOS\space}%
\providecommand \EOS [0]{\spacefactor3000\relax}%
\providecommand \BibitemShut  [1]{\csname bibitem#1\endcsname}%
\let\auto@bib@innerbib\@empty
%</preamble>
\bibitem [{\citenamefont {Kjaergaard}\ \emph {et~al.}(2020)\citenamefont
  {Kjaergaard}, \citenamefont {Schwartz}, \citenamefont {Braumüller},
  \citenamefont {Krantz}, \citenamefont {Wang}, \citenamefont {Gustavsson},\
  and\ \citenamefont {Oliver}}]{kjaergaard2019superconducting}%
  \BibitemOpen
  \bibfield  {author} {\bibinfo {author} {\bibfnamefont {M.}~\bibnamefont
  {Kjaergaard}}, \bibinfo {author} {\bibfnamefont {M.~E.}\ \bibnamefont
  {Schwartz}}, \bibinfo {author} {\bibfnamefont {J.}~\bibnamefont
  {Braumüller}}, \bibinfo {author} {\bibfnamefont {P.}~\bibnamefont {Krantz}},
  \bibinfo {author} {\bibfnamefont {J.~I.-J.}\ \bibnamefont {Wang}}, \bibinfo
  {author} {\bibfnamefont {S.}~\bibnamefont {Gustavsson}}, \ and\ \bibinfo
  {author} {\bibfnamefont {W.~D.}\ \bibnamefont {Oliver}},\ }\href {\doibase
  10.1146/annurev-conmatphys-031119-050605} {\bibfield  {journal} {\bibinfo
  {journal} {Annu. Rev. Condens. Matter Phys.}\ }\textbf {\bibinfo {volume}
  {11}},\ \bibinfo {pages} {369} (\bibinfo {year} {2020})},\ \Eprint
  {http://arxiv.org/abs/https://doi.org/10.1146/annurev-conmatphys-031119-050605}
  {https://doi.org/10.1146/annurev-conmatphys-031119-050605} \BibitemShut
  {NoStop}%
\bibitem [{\citenamefont {Orell}\ \emph {et~al.}(2019)\citenamefont {Orell},
  \citenamefont {Michailidis}, \citenamefont {Serbyn},\ and\ \citenamefont
  {Silveri}}]{orell2019probing}%
  \BibitemOpen
  \bibfield  {author} {\bibinfo {author} {\bibfnamefont {T.}~\bibnamefont
  {Orell}}, \bibinfo {author} {\bibfnamefont {A.~A.}\ \bibnamefont
  {Michailidis}}, \bibinfo {author} {\bibfnamefont {M.}~\bibnamefont {Serbyn}},
  \ and\ \bibinfo {author} {\bibfnamefont {M.}~\bibnamefont {Silveri}},\ }\href
  {\doibase 10.1103/PhysRevB.100.134504} {\bibfield  {journal} {\bibinfo
  {journal} {Phys. Rev. B}\ }\textbf {\bibinfo {volume} {100}},\ \bibinfo
  {pages} {134504} (\bibinfo {year} {2019})}\BibitemShut {NoStop}%
\bibitem [{\citenamefont {Yanay}\ \emph {et~al.}(2020)\citenamefont {Yanay},
  \citenamefont {Braum{\"u}ller}, \citenamefont {Gustavsson}, \citenamefont
  {Oliver},\ and\ \citenamefont {Tahan}}]{yanay2020two}%
  \BibitemOpen
  \bibfield  {author} {\bibinfo {author} {\bibfnamefont {Y.}~\bibnamefont
  {Yanay}}, \bibinfo {author} {\bibfnamefont {J.}~\bibnamefont
  {Braum{\"u}ller}}, \bibinfo {author} {\bibfnamefont {S.}~\bibnamefont
  {Gustavsson}}, \bibinfo {author} {\bibfnamefont {W.~D.}\ \bibnamefont
  {Oliver}}, \ and\ \bibinfo {author} {\bibfnamefont {C.}~\bibnamefont
  {Tahan}},\ }\href@noop {} {\bibfield  {journal} {\bibinfo  {journal} {npj
  Quantum Inf.}\ }\textbf {\bibinfo {volume} {6}},\ \bibinfo {pages} {1}
  (\bibinfo {year} {2020})}\BibitemShut {NoStop}%
\bibitem [{\citenamefont {Hacohen-Gourgy}\ \emph {et~al.}(2015)\citenamefont
  {Hacohen-Gourgy}, \citenamefont {Ramasesh}, \citenamefont {De~Grandi},
  \citenamefont {Siddiqi},\ and\ \citenamefont {Girvin}}]{hacohen2015cooling}%
  \BibitemOpen
  \bibfield  {author} {\bibinfo {author} {\bibfnamefont {S.}~\bibnamefont
  {Hacohen-Gourgy}}, \bibinfo {author} {\bibfnamefont {V.~V.}\ \bibnamefont
  {Ramasesh}}, \bibinfo {author} {\bibfnamefont {C.}~\bibnamefont {De~Grandi}},
  \bibinfo {author} {\bibfnamefont {I.}~\bibnamefont {Siddiqi}}, \ and\
  \bibinfo {author} {\bibfnamefont {S.~M.}\ \bibnamefont {Girvin}},\
  }\href@noop {} {\bibfield  {journal} {\bibinfo  {journal} {Phys. Rev.
  letters}\ }\textbf {\bibinfo {volume} {115}},\ \bibinfo {pages} {240501}
  (\bibinfo {year} {2015})}\BibitemShut {NoStop}%
\bibitem [{\citenamefont {Ma}\ \emph {et~al.}(2019)\citenamefont {Ma},
  \citenamefont {Saxberg}, \citenamefont {Owens}, \citenamefont {Leung},
  \citenamefont {Lu}, \citenamefont {Simon},\ and\ \citenamefont
  {Schuster}}]{ma2019dissipatively}%
  \BibitemOpen
  \bibfield  {author} {\bibinfo {author} {\bibfnamefont {R.}~\bibnamefont
  {Ma}}, \bibinfo {author} {\bibfnamefont {B.}~\bibnamefont {Saxberg}},
  \bibinfo {author} {\bibfnamefont {C.}~\bibnamefont {Owens}}, \bibinfo
  {author} {\bibfnamefont {N.}~\bibnamefont {Leung}}, \bibinfo {author}
  {\bibfnamefont {Y.}~\bibnamefont {Lu}}, \bibinfo {author} {\bibfnamefont
  {J.}~\bibnamefont {Simon}}, \ and\ \bibinfo {author} {\bibfnamefont {D.~I.}\
  \bibnamefont {Schuster}},\ }\href@noop {} {\bibfield  {journal} {\bibinfo
  {journal} {Nature}\ }\textbf {\bibinfo {volume} {566}},\ \bibinfo {pages}
  {51} (\bibinfo {year} {2019})}\BibitemShut {NoStop}%
\bibitem [{\citenamefont {Roushan}\ \emph {et~al.}(2017)\citenamefont
  {Roushan}, \citenamefont {Neill}, \citenamefont {Tangpanitanon},
  \citenamefont {Bastidas}, \citenamefont {Megrant}, \citenamefont {Barends},
  \citenamefont {Chen}, \citenamefont {Chen}, \citenamefont {Chiaro},
  \citenamefont {Dunsworth} \emph {et~al.}}]{roushan2017spectroscopic}%
  \BibitemOpen
  \bibfield  {author} {\bibinfo {author} {\bibfnamefont {P.}~\bibnamefont
  {Roushan}}, \bibinfo {author} {\bibfnamefont {C.}~\bibnamefont {Neill}},
  \bibinfo {author} {\bibfnamefont {J.}~\bibnamefont {Tangpanitanon}}, \bibinfo
  {author} {\bibfnamefont {V.}~\bibnamefont {Bastidas}}, \bibinfo {author}
  {\bibfnamefont {A.}~\bibnamefont {Megrant}}, \bibinfo {author} {\bibfnamefont
  {R.}~\bibnamefont {Barends}}, \bibinfo {author} {\bibfnamefont
  {Y.}~\bibnamefont {Chen}}, \bibinfo {author} {\bibfnamefont {Z.}~\bibnamefont
  {Chen}}, \bibinfo {author} {\bibfnamefont {B.}~\bibnamefont {Chiaro}},
  \bibinfo {author} {\bibfnamefont {A.}~\bibnamefont {Dunsworth}},  \emph
  {et~al.},\ }\href@noop {} {\bibfield  {journal} {\bibinfo  {journal}
  {Science}\ }\textbf {\bibinfo {volume} {358}},\ \bibinfo {pages} {1175}
  (\bibinfo {year} {2017})}\BibitemShut {NoStop}%
\bibitem [{\citenamefont {Chiaro}\ \emph {et~al.}(2019)\citenamefont {Chiaro},
  \citenamefont {Neill}, \citenamefont {Bohrdt}, \citenamefont {Filippone},
  \citenamefont {Arute}, \citenamefont {Arya}, \citenamefont {Babbush},
  \citenamefont {Bacon}, \citenamefont {Bardin}, \citenamefont {Barends} \emph
  {et~al.}}]{chiaro2019growth}%
  \BibitemOpen
  \bibfield  {author} {\bibinfo {author} {\bibfnamefont {B.}~\bibnamefont
  {Chiaro}}, \bibinfo {author} {\bibfnamefont {C.}~\bibnamefont {Neill}},
  \bibinfo {author} {\bibfnamefont {A.}~\bibnamefont {Bohrdt}}, \bibinfo
  {author} {\bibfnamefont {M.}~\bibnamefont {Filippone}}, \bibinfo {author}
  {\bibfnamefont {F.}~\bibnamefont {Arute}}, \bibinfo {author} {\bibfnamefont
  {K.}~\bibnamefont {Arya}}, \bibinfo {author} {\bibfnamefont {R.}~\bibnamefont
  {Babbush}}, \bibinfo {author} {\bibfnamefont {D.}~\bibnamefont {Bacon}},
  \bibinfo {author} {\bibfnamefont {J.}~\bibnamefont {Bardin}}, \bibinfo
  {author} {\bibfnamefont {R.}~\bibnamefont {Barends}},  \emph {et~al.},\
  }\href@noop {} {\bibfield  {journal} {\bibinfo  {journal} {arXiv:1910.06024}\
  } (\bibinfo {year} {2019})}\BibitemShut {NoStop}%
\bibitem [{\citenamefont {Yan}\ \emph {et~al.}(2019)\citenamefont {Yan},
  \citenamefont {Zhang}, \citenamefont {Gong}, \citenamefont {Wu},
  \citenamefont {Zheng}, \citenamefont {Li}, \citenamefont {Wang},
  \citenamefont {Liang}, \citenamefont {Lin}, \citenamefont {Xu}, \citenamefont
  {Guo}, \citenamefont {Sun}, \citenamefont {Peng}, \citenamefont {Xia},
  \citenamefont {Deng}, \citenamefont {Rong}, \citenamefont {You},
  \citenamefont {Nori}, \citenamefont {Fan}, \citenamefont {Zhu},\ and\
  \citenamefont {Pan}}]{Yan2019}%
  \BibitemOpen
  \bibfield  {author} {\bibinfo {author} {\bibfnamefont {Z.}~\bibnamefont
  {Yan}}, \bibinfo {author} {\bibfnamefont {Y.~R.}\ \bibnamefont {Zhang}},
  \bibinfo {author} {\bibfnamefont {M.}~\bibnamefont {Gong}}, \bibinfo {author}
  {\bibfnamefont {Y.}~\bibnamefont {Wu}}, \bibinfo {author} {\bibfnamefont
  {Y.}~\bibnamefont {Zheng}}, \bibinfo {author} {\bibfnamefont
  {S.}~\bibnamefont {Li}}, \bibinfo {author} {\bibfnamefont {C.}~\bibnamefont
  {Wang}}, \bibinfo {author} {\bibfnamefont {F.}~\bibnamefont {Liang}},
  \bibinfo {author} {\bibfnamefont {J.}~\bibnamefont {Lin}}, \bibinfo {author}
  {\bibfnamefont {Y.}~\bibnamefont {Xu}}, \bibinfo {author} {\bibfnamefont
  {C.}~\bibnamefont {Guo}}, \bibinfo {author} {\bibfnamefont {L.}~\bibnamefont
  {Sun}}, \bibinfo {author} {\bibfnamefont {C.~Z.}\ \bibnamefont {Peng}},
  \bibinfo {author} {\bibfnamefont {K.}~\bibnamefont {Xia}}, \bibinfo {author}
  {\bibfnamefont {H.}~\bibnamefont {Deng}}, \bibinfo {author} {\bibfnamefont
  {H.}~\bibnamefont {Rong}}, \bibinfo {author} {\bibfnamefont {J.~Q.}\
  \bibnamefont {You}}, \bibinfo {author} {\bibfnamefont {F.}~\bibnamefont
  {Nori}}, \bibinfo {author} {\bibfnamefont {H.}~\bibnamefont {Fan}}, \bibinfo
  {author} {\bibfnamefont {X.}~\bibnamefont {Zhu}}, \ and\ \bibinfo {author}
  {\bibfnamefont {J.~W.}\ \bibnamefont {Pan}},\ }\href {\doibase
  10.1126/science.aaw1611} {\bibfield  {journal} {\bibinfo  {journal}
  {Science}\ }\textbf {\bibinfo {volume} {756}},\ \bibinfo {pages} {753}
  (\bibinfo {year} {2019})}\BibitemShut {NoStop}%
\bibitem [{\citenamefont {Ye}\ \emph {et~al.}(2019)\citenamefont {Ye},
  \citenamefont {Ge}, \citenamefont {Wu}, \citenamefont {Wang}, \citenamefont
  {Gong}, \citenamefont {Zhang}, \citenamefont {Zhu}, \citenamefont {Yang},
  \citenamefont {Li}, \citenamefont {Liang}, \citenamefont {Lin}, \citenamefont
  {Xu}, \citenamefont {Guo}, \citenamefont {Sun}, \citenamefont {Cheng},
  \citenamefont {Ma}, \citenamefont {Meng}, \citenamefont {Deng}, \citenamefont
  {Rong}, \citenamefont {Lu}, \citenamefont {Peng}, \citenamefont {Fan},
  \citenamefont {Zhu},\ and\ \citenamefont {Pan}}]{Ye2019}%
  \BibitemOpen
  \bibfield  {author} {\bibinfo {author} {\bibfnamefont {Y.}~\bibnamefont
  {Ye}}, \bibinfo {author} {\bibfnamefont {Z.~Y.}\ \bibnamefont {Ge}}, \bibinfo
  {author} {\bibfnamefont {Y.}~\bibnamefont {Wu}}, \bibinfo {author}
  {\bibfnamefont {S.}~\bibnamefont {Wang}}, \bibinfo {author} {\bibfnamefont
  {M.}~\bibnamefont {Gong}}, \bibinfo {author} {\bibfnamefont {Y.~R.}\
  \bibnamefont {Zhang}}, \bibinfo {author} {\bibfnamefont {Q.}~\bibnamefont
  {Zhu}}, \bibinfo {author} {\bibfnamefont {R.}~\bibnamefont {Yang}}, \bibinfo
  {author} {\bibfnamefont {S.}~\bibnamefont {Li}}, \bibinfo {author}
  {\bibfnamefont {F.}~\bibnamefont {Liang}}, \bibinfo {author} {\bibfnamefont
  {J.}~\bibnamefont {Lin}}, \bibinfo {author} {\bibfnamefont {Y.}~\bibnamefont
  {Xu}}, \bibinfo {author} {\bibfnamefont {C.}~\bibnamefont {Guo}}, \bibinfo
  {author} {\bibfnamefont {L.}~\bibnamefont {Sun}}, \bibinfo {author}
  {\bibfnamefont {C.}~\bibnamefont {Cheng}}, \bibinfo {author} {\bibfnamefont
  {N.}~\bibnamefont {Ma}}, \bibinfo {author} {\bibfnamefont {Z.~Y.}\
  \bibnamefont {Meng}}, \bibinfo {author} {\bibfnamefont {H.}~\bibnamefont
  {Deng}}, \bibinfo {author} {\bibfnamefont {H.}~\bibnamefont {Rong}}, \bibinfo
  {author} {\bibfnamefont {C.~Y.}\ \bibnamefont {Lu}}, \bibinfo {author}
  {\bibfnamefont {C.~Z.}\ \bibnamefont {Peng}}, \bibinfo {author}
  {\bibfnamefont {H.}~\bibnamefont {Fan}}, \bibinfo {author} {\bibfnamefont
  {X.}~\bibnamefont {Zhu}}, \ and\ \bibinfo {author} {\bibfnamefont {J.~W.}\
  \bibnamefont {Pan}},\ }\href {\doibase 10.1103/PhysRevLett.123.050502}
  {\bibfield  {journal} {\bibinfo  {journal} {Phys. Rev. Letters}\ }\textbf
  {\bibinfo {volume} {123}},\ \bibinfo {pages} {1} (\bibinfo {year}
  {2019})}\BibitemShut {NoStop}%
\bibitem [{\citenamefont {Goldman}\ and\ \citenamefont
  {Dalibard}(2014)}]{Goldman2014}%
  \BibitemOpen
  \bibfield  {author} {\bibinfo {author} {\bibfnamefont {N.}~\bibnamefont
  {Goldman}}\ and\ \bibinfo {author} {\bibfnamefont {J.}~\bibnamefont
  {Dalibard}},\ }\href {\doibase 10.1103/PhysRevX.4.031027} {\bibfield
  {journal} {\bibinfo  {journal} {Phys. Rev. X}\ }\textbf {\bibinfo {volume}
  {4}},\ \bibinfo {pages} {1} (\bibinfo {year} {2014})}\BibitemShut {NoStop}%
\bibitem [{\citenamefont {Eisert}\ \emph {et~al.}(2015)\citenamefont {Eisert},
  \citenamefont {Friesdorf},\ and\ \citenamefont
  {Gogolin}}]{eisert2015quantum}%
  \BibitemOpen
  \bibfield  {author} {\bibinfo {author} {\bibfnamefont {J.}~\bibnamefont
  {Eisert}}, \bibinfo {author} {\bibfnamefont {M.}~\bibnamefont {Friesdorf}}, \
  and\ \bibinfo {author} {\bibfnamefont {C.}~\bibnamefont {Gogolin}},\
  }\href@noop {} {\bibfield  {journal} {\bibinfo  {journal} {Nat. Phys.}\
  }\textbf {\bibinfo {volume} {11}},\ \bibinfo {pages} {124} (\bibinfo {year}
  {2015})}\BibitemShut {NoStop}%
\bibitem [{\citenamefont {Zippilli}\ \emph {et~al.}(2015)\citenamefont
  {Zippilli}, \citenamefont {Grajcar}, \citenamefont {Il'Ichev},\ and\
  \citenamefont {Illuminati}}]{Zippilli2015}%
  \BibitemOpen
  \bibfield  {author} {\bibinfo {author} {\bibfnamefont {S.}~\bibnamefont
  {Zippilli}}, \bibinfo {author} {\bibfnamefont {M.}~\bibnamefont {Grajcar}},
  \bibinfo {author} {\bibfnamefont {E.}~\bibnamefont {Il'Ichev}}, \ and\
  \bibinfo {author} {\bibfnamefont {F.}~\bibnamefont {Illuminati}},\ }\href
  {\doibase 10.1103/PhysRevA.91.022315} {\bibfield  {journal} {\bibinfo
  {journal} {Phys. Rev. A}\ }\textbf {\bibinfo {volume} {91}},\ \bibinfo
  {pages} {1} (\bibinfo {year} {2015})}\BibitemShut {NoStop}%
\bibitem [{\citenamefont {Kyriienko}\ and\ \citenamefont
  {S{\o}rensen}(2018)}]{kyriienko2018floquet}%
  \BibitemOpen
  \bibfield  {author} {\bibinfo {author} {\bibfnamefont {O.}~\bibnamefont
  {Kyriienko}}\ and\ \bibinfo {author} {\bibfnamefont {A.~S.}\ \bibnamefont
  {S{\o}rensen}},\ }\href@noop {} {\bibfield  {journal} {\bibinfo  {journal}
  {Phys. Rev. Applied}\ }\textbf {\bibinfo {volume} {9}},\ \bibinfo {pages}
  {064029} (\bibinfo {year} {2018})}\BibitemShut {NoStop}%
\bibitem [{\citenamefont {Franca}\ \emph {et~al.}(2020)\citenamefont {Franca},
  \citenamefont {Hassler},\ and\ \citenamefont {Fulga}}]{franca2020simulating}%
  \BibitemOpen
  \bibfield  {author} {\bibinfo {author} {\bibfnamefont {S.}~\bibnamefont
  {Franca}}, \bibinfo {author} {\bibfnamefont {F.}~\bibnamefont {Hassler}}, \
  and\ \bibinfo {author} {\bibfnamefont {I.~C.}\ \bibnamefont {Fulga}},\
  }\href@noop {} {\bibfield  {journal} {\bibinfo  {journal} {arXiv:2001.08217}\
  } (\bibinfo {year} {2020})}\BibitemShut {NoStop}%
\bibitem [{\citenamefont {Tangpanitanon}\ \emph {et~al.}(2019)\citenamefont
  {Tangpanitanon}, \citenamefont {Thanasilp}, \citenamefont {Lemonde},\ and\
  \citenamefont {Angelakis}}]{tangpanitanon2019quantum}%
  \BibitemOpen
  \bibfield  {author} {\bibinfo {author} {\bibfnamefont {J.}~\bibnamefont
  {Tangpanitanon}}, \bibinfo {author} {\bibfnamefont {S.}~\bibnamefont
  {Thanasilp}}, \bibinfo {author} {\bibfnamefont {M.-A.}\ \bibnamefont
  {Lemonde}}, \ and\ \bibinfo {author} {\bibfnamefont {D.~G.}\ \bibnamefont
  {Angelakis}},\ }\href@noop {} {\bibfield  {journal} {\bibinfo  {journal}
  {arXiv:1906.03860}\ } (\bibinfo {year} {2019})}\BibitemShut {NoStop}%
\bibitem [{\citenamefont {Zagoskin}\ \emph {et~al.}(2016)\citenamefont
  {Zagoskin}, \citenamefont {Felbacq},\ and\ \citenamefont
  {Rousseau}}]{Zagoskin2016}%
  \BibitemOpen
  \bibfield  {author} {\bibinfo {author} {\bibfnamefont {A.~M.}\ \bibnamefont
  {Zagoskin}}, \bibinfo {author} {\bibfnamefont {D.}~\bibnamefont {Felbacq}}, \
  and\ \bibinfo {author} {\bibfnamefont {E.}~\bibnamefont {Rousseau}},\ }\href
  {\doibase 10.1140/epjqt/s40507-016-0040-x} {\enquote {\bibinfo {title}
  {{Quantum metamaterials in the microwave and optical ranges}},}\ } (\bibinfo
  {year} {2016})\BibitemShut {NoStop}%
\bibitem [{\citenamefont {Viehmann}\ \emph {et~al.}(2013)\citenamefont
  {Viehmann}, \citenamefont {von Delft},\ and\ \citenamefont
  {Marquardt}}]{viehmann2013observing}%
  \BibitemOpen
  \bibfield  {author} {\bibinfo {author} {\bibfnamefont {O.}~\bibnamefont
  {Viehmann}}, \bibinfo {author} {\bibfnamefont {J.}~\bibnamefont {von Delft}},
  \ and\ \bibinfo {author} {\bibfnamefont {F.}~\bibnamefont {Marquardt}},\
  }\href@noop {} {\bibfield  {journal} {\bibinfo  {journal} {Phys. Rev.
  letters}\ }\textbf {\bibinfo {volume} {110}},\ \bibinfo {pages} {030601}
  (\bibinfo {year} {2013})}\BibitemShut {NoStop}%
\bibitem [{\citenamefont {Greenberg}\ and\ \citenamefont
  {Shtygashev}(2015)}]{Greenberg2015}%
  \BibitemOpen
  \bibfield  {author} {\bibinfo {author} {\bibfnamefont {Y.~S.}\ \bibnamefont
  {Greenberg}}\ and\ \bibinfo {author} {\bibfnamefont {A.~A.}\ \bibnamefont
  {Shtygashev}},\ }\href {\doibase 10.1103/PhysRevA.92.063835} {\bibfield
  {journal} {\bibinfo  {journal} {Phys. Rev. A}\ }\textbf {\bibinfo {volume}
  {92}},\ \bibinfo {pages} {1} (\bibinfo {year} {2015})}\BibitemShut {NoStop}%
\bibitem [{\citenamefont {Fistul}\ and\ \citenamefont
  {Iontsev}(2019)}]{Fistul2019}%
  \BibitemOpen
  \bibfield  {author} {\bibinfo {author} {\bibfnamefont {M.~V.}\ \bibnamefont
  {Fistul}}\ and\ \bibinfo {author} {\bibfnamefont {M.~A.}\ \bibnamefont
  {Iontsev}},\ }\href {\doibase 10.1103/PhysRevA.100.023844} {\bibfield
  {journal} {\bibinfo  {journal} {Phys. Rev. A}\ }\textbf {\bibinfo {volume}
  {100}},\ \bibinfo {pages} {1} (\bibinfo {year} {2019})}\BibitemShut {NoStop}%
\bibitem [{\citenamefont {Biella}\ \emph {et~al.}(2015)\citenamefont {Biella},
  \citenamefont {Mazza}, \citenamefont {Carusotto}, \citenamefont {Rossini},\
  and\ \citenamefont {Fazio}}]{Biella2015}%
  \BibitemOpen
  \bibfield  {author} {\bibinfo {author} {\bibfnamefont {A.}~\bibnamefont
  {Biella}}, \bibinfo {author} {\bibfnamefont {L.}~\bibnamefont {Mazza}},
  \bibinfo {author} {\bibfnamefont {I.}~\bibnamefont {Carusotto}}, \bibinfo
  {author} {\bibfnamefont {D.}~\bibnamefont {Rossini}}, \ and\ \bibinfo
  {author} {\bibfnamefont {R.}~\bibnamefont {Fazio}},\ }\href {\doibase
  10.1103/PhysRevA.91.053815} {\bibfield  {journal} {\bibinfo  {journal} {Phys.
  Rev. A}\ }\textbf {\bibinfo {volume} {91}},\ \bibinfo {pages} {1} (\bibinfo
  {year} {2015})}\BibitemShut {NoStop}%
\bibitem [{\citenamefont {Roberts}\ and\ \citenamefont
  {Clerk}(2020)}]{roberts2020driven}%
  \BibitemOpen
  \bibfield  {author} {\bibinfo {author} {\bibfnamefont {D.}~\bibnamefont
  {Roberts}}\ and\ \bibinfo {author} {\bibfnamefont {A.~A.}\ \bibnamefont
  {Clerk}},\ }\href@noop {} {\bibfield  {journal} {\bibinfo  {journal} {Phys.
  Rev. X}\ }\textbf {\bibinfo {volume} {10}},\ \bibinfo {pages} {021022}
  (\bibinfo {year} {2020})}\BibitemShut {NoStop}%
\bibitem [{\citenamefont {Collodo}\ \emph {et~al.}(2019)\citenamefont
  {Collodo}, \citenamefont {Poto{\v{c}}nik}, \citenamefont {Gasparinetti},
  \citenamefont {Besse}, \citenamefont {Pechal}, \citenamefont {Sameti},
  \citenamefont {Hartmann}, \citenamefont {Wallraff},\ and\ \citenamefont
  {Eichler}}]{collodo2019observation}%
  \BibitemOpen
  \bibfield  {author} {\bibinfo {author} {\bibfnamefont {M.~C.}\ \bibnamefont
  {Collodo}}, \bibinfo {author} {\bibfnamefont {A.}~\bibnamefont
  {Poto{\v{c}}nik}}, \bibinfo {author} {\bibfnamefont {S.}~\bibnamefont
  {Gasparinetti}}, \bibinfo {author} {\bibfnamefont {J.-C.}\ \bibnamefont
  {Besse}}, \bibinfo {author} {\bibfnamefont {M.}~\bibnamefont {Pechal}},
  \bibinfo {author} {\bibfnamefont {M.}~\bibnamefont {Sameti}}, \bibinfo
  {author} {\bibfnamefont {M.~J.}\ \bibnamefont {Hartmann}}, \bibinfo {author}
  {\bibfnamefont {A.}~\bibnamefont {Wallraff}}, \ and\ \bibinfo {author}
  {\bibfnamefont {C.}~\bibnamefont {Eichler}},\ }\href@noop {} {\bibfield
  {journal} {\bibinfo  {journal} {Phys. Rev. letters}\ }\textbf {\bibinfo
  {volume} {122}},\ \bibinfo {pages} {183601} (\bibinfo {year}
  {2019})}\BibitemShut {NoStop}%
\bibitem [{\citenamefont {Tiwari}\ \emph {et~al.}(2020)\citenamefont {Tiwari},
  \citenamefont {Roy},\ and\ \citenamefont {Singh}}]{tiwari2020interplay}%
  \BibitemOpen
  \bibfield  {author} {\bibinfo {author} {\bibfnamefont {T.}~\bibnamefont
  {Tiwari}}, \bibinfo {author} {\bibfnamefont {D.}~\bibnamefont {Roy}}, \ and\
  \bibinfo {author} {\bibfnamefont {R.}~\bibnamefont {Singh}},\ }\href@noop {}
  {\bibfield  {journal} {\bibinfo  {journal} {arXiv:2010.14935}\ } (\bibinfo
  {year} {2020})}\BibitemShut {NoStop}%
\bibitem [{\citenamefont {Di~Paolo}\ \emph {et~al.}(2019)\citenamefont
  {Di~Paolo}, \citenamefont {Baker}, \citenamefont {Foley}, \citenamefont
  {S{\'e}n{\'e}chal},\ and\ \citenamefont {Blais}}]{di2019efficient}%
  \BibitemOpen
  \bibfield  {author} {\bibinfo {author} {\bibfnamefont {A.}~\bibnamefont
  {Di~Paolo}}, \bibinfo {author} {\bibfnamefont {T.~E.}\ \bibnamefont {Baker}},
  \bibinfo {author} {\bibfnamefont {A.}~\bibnamefont {Foley}}, \bibinfo
  {author} {\bibfnamefont {D.}~\bibnamefont {S{\'e}n{\'e}chal}}, \ and\
  \bibinfo {author} {\bibfnamefont {A.}~\bibnamefont {Blais}},\ }\href@noop {}
  {\bibfield  {journal} {\bibinfo  {journal} {arXiv:1912.01018}\ } (\bibinfo
  {year} {2019})}\BibitemShut {NoStop}%
\bibitem [{\citenamefont {Egorova}\ \emph {et~al.}(2020)\citenamefont
  {Egorova}, \citenamefont {Fedorov}, \citenamefont {Tsitsilin}, \citenamefont
  {Besedin},\ and\ \citenamefont {Ustinov}}]{egorova2020analog}%
  \BibitemOpen
  \bibfield  {author} {\bibinfo {author} {\bibfnamefont {E.}~\bibnamefont
  {Egorova}}, \bibinfo {author} {\bibfnamefont {G.}~\bibnamefont {Fedorov}},
  \bibinfo {author} {\bibfnamefont {I.}~\bibnamefont {Tsitsilin}}, \bibinfo
  {author} {\bibfnamefont {I.}~\bibnamefont {Besedin}}, \ and\ \bibinfo
  {author} {\bibfnamefont {A.}~\bibnamefont {Ustinov}},\ }in\ \href@noop {}
  {\emph {\bibinfo {booktitle} {AIP Conf. Proc.}}},\ Vol.\ \bibinfo {volume}
  {2241}\ (\bibinfo {year} {2020})\ p.\ \bibinfo {pages} {020013}\BibitemShut
  {NoStop}%
\bibitem [{\citenamefont {Fedorov}\ \emph {et~al.}(2020)\citenamefont
  {Fedorov}, \citenamefont {Yursa}, \citenamefont {Efimov}, \citenamefont
  {Shiianov}, \citenamefont {Dmitriev}, \citenamefont {Rodionov}, \citenamefont
  {Dobronosova}, \citenamefont {Moskalev}, \citenamefont {Pishchimova},
  \citenamefont {Malevannaya},\ and\ \citenamefont
  {Astafiev}}]{PhysRevA.102.013707}%
  \BibitemOpen
  \bibfield  {author} {\bibinfo {author} {\bibfnamefont {G.~P.}\ \bibnamefont
  {Fedorov}}, \bibinfo {author} {\bibfnamefont {V.~B.}\ \bibnamefont {Yursa}},
  \bibinfo {author} {\bibfnamefont {A.~E.}\ \bibnamefont {Efimov}}, \bibinfo
  {author} {\bibfnamefont {K.~I.}\ \bibnamefont {Shiianov}}, \bibinfo {author}
  {\bibfnamefont {A.~Y.}\ \bibnamefont {Dmitriev}}, \bibinfo {author}
  {\bibfnamefont {I.~A.}\ \bibnamefont {Rodionov}}, \bibinfo {author}
  {\bibfnamefont {A.~A.}\ \bibnamefont {Dobronosova}}, \bibinfo {author}
  {\bibfnamefont {D.~O.}\ \bibnamefont {Moskalev}}, \bibinfo {author}
  {\bibfnamefont {A.~A.}\ \bibnamefont {Pishchimova}}, \bibinfo {author}
  {\bibfnamefont {E.~I.}\ \bibnamefont {Malevannaya}}, \ and\ \bibinfo {author}
  {\bibfnamefont {O.~V.}\ \bibnamefont {Astafiev}},\ }\href {\doibase
  10.1103/PhysRevA.102.013707} {\bibfield  {journal} {\bibinfo  {journal}
  {Phys. Rev. A}\ }\textbf {\bibinfo {volume} {102}},\ \bibinfo {pages}
  {013707} (\bibinfo {year} {2020})}\BibitemShut {NoStop}%
\bibitem [{\citenamefont {Yurke}\ and\ \citenamefont
  {Denker}(1984)}]{yurke1984quantum}%
  \BibitemOpen
  \bibfield  {author} {\bibinfo {author} {\bibfnamefont {B.}~\bibnamefont
  {Yurke}}\ and\ \bibinfo {author} {\bibfnamefont {J.~S.}\ \bibnamefont
  {Denker}},\ }\href@noop {} {\bibfield  {journal} {\bibinfo  {journal} {Phys.
  Rev. A}\ }\textbf {\bibinfo {volume} {29}},\ \bibinfo {pages} {1419}
  (\bibinfo {year} {1984})}\BibitemShut {NoStop}%
\bibitem [{\citenamefont {Gardiner}\ and\ \citenamefont
  {Collett}(1985)}]{gardiner1985input}%
  \BibitemOpen
  \bibfield  {author} {\bibinfo {author} {\bibfnamefont {C.~W.}\ \bibnamefont
  {Gardiner}}\ and\ \bibinfo {author} {\bibfnamefont {M.~J.}\ \bibnamefont
  {Collett}},\ }\href@noop {} {\bibfield  {journal} {\bibinfo  {journal} {Phys.
  Rev. A}\ }\textbf {\bibinfo {volume} {31}},\ \bibinfo {pages} {3761}
  (\bibinfo {year} {1985})}\BibitemShut {NoStop}%
\bibitem [{\citenamefont {Mirhosseini}\ \emph {et~al.}(2019)\citenamefont
  {Mirhosseini}, \citenamefont {Kim}, \citenamefont {Zhang}, \citenamefont
  {Sipahigil}, \citenamefont {Dieterle}, \citenamefont {Keller}, \citenamefont
  {Asenjo-Garcia}, \citenamefont {Chang},\ and\ \citenamefont
  {Painter}}]{mirhosseini2019cavity}%
  \BibitemOpen
  \bibfield  {author} {\bibinfo {author} {\bibfnamefont {M.}~\bibnamefont
  {Mirhosseini}}, \bibinfo {author} {\bibfnamefont {E.}~\bibnamefont {Kim}},
  \bibinfo {author} {\bibfnamefont {X.}~\bibnamefont {Zhang}}, \bibinfo
  {author} {\bibfnamefont {A.}~\bibnamefont {Sipahigil}}, \bibinfo {author}
  {\bibfnamefont {P.~B.}\ \bibnamefont {Dieterle}}, \bibinfo {author}
  {\bibfnamefont {A.~J.}\ \bibnamefont {Keller}}, \bibinfo {author}
  {\bibfnamefont {A.}~\bibnamefont {Asenjo-Garcia}}, \bibinfo {author}
  {\bibfnamefont {D.~E.}\ \bibnamefont {Chang}}, \ and\ \bibinfo {author}
  {\bibfnamefont {O.}~\bibnamefont {Painter}},\ }\href@noop {} {\bibfield
  {journal} {\bibinfo  {journal} {Nature}\ }\textbf {\bibinfo {volume} {569}},\
  \bibinfo {pages} {692} (\bibinfo {year} {2019})}\BibitemShut {NoStop}%
\bibitem [{\citenamefont {Astafiev}\ \emph {et~al.}(2010)\citenamefont
  {Astafiev}, \citenamefont {Zagoskin}, \citenamefont {Abdumalikov},
  \citenamefont {Pashkin}, \citenamefont {Yamamoto}, \citenamefont {Inomata},
  \citenamefont {Nakamura},\ and\ \citenamefont
  {Tsai}}]{astafiev2010resonance}%
  \BibitemOpen
  \bibfield  {author} {\bibinfo {author} {\bibfnamefont {O.}~\bibnamefont
  {Astafiev}}, \bibinfo {author} {\bibfnamefont {A.~M.}\ \bibnamefont
  {Zagoskin}}, \bibinfo {author} {\bibfnamefont {A.}~\bibnamefont
  {Abdumalikov}}, \bibinfo {author} {\bibfnamefont {Y.~A.}\ \bibnamefont
  {Pashkin}}, \bibinfo {author} {\bibfnamefont {T.}~\bibnamefont {Yamamoto}},
  \bibinfo {author} {\bibfnamefont {K.}~\bibnamefont {Inomata}}, \bibinfo
  {author} {\bibfnamefont {Y.}~\bibnamefont {Nakamura}}, \ and\ \bibinfo
  {author} {\bibfnamefont {J.~S.}\ \bibnamefont {Tsai}},\ }\href@noop {}
  {\bibfield  {journal} {\bibinfo  {journal} {Science}\ }\textbf {\bibinfo
  {volume} {327}},\ \bibinfo {pages} {840} (\bibinfo {year}
  {2010})}\BibitemShut {NoStop}%
\bibitem [{\citenamefont {Birnbaum}\ \emph {et~al.}(2005)\citenamefont
  {Birnbaum}, \citenamefont {Boca}, \citenamefont {Miller}, \citenamefont
  {Boozer}, \citenamefont {Northup},\ and\ \citenamefont
  {Kimble}}]{birnbaum2005photon}%
  \BibitemOpen
  \bibfield  {author} {\bibinfo {author} {\bibfnamefont {K.~M.}\ \bibnamefont
  {Birnbaum}}, \bibinfo {author} {\bibfnamefont {A.}~\bibnamefont {Boca}},
  \bibinfo {author} {\bibfnamefont {R.}~\bibnamefont {Miller}}, \bibinfo
  {author} {\bibfnamefont {A.~D.}\ \bibnamefont {Boozer}}, \bibinfo {author}
  {\bibfnamefont {T.~E.}\ \bibnamefont {Northup}}, \ and\ \bibinfo {author}
  {\bibfnamefont {H.~J.}\ \bibnamefont {Kimble}},\ }\href@noop {} {\bibfield
  {journal} {\bibinfo  {journal} {Nature}\ }\textbf {\bibinfo {volume} {436}},\
  \bibinfo {pages} {87} (\bibinfo {year} {2005})}\BibitemShut {NoStop}%
\bibitem [{\citenamefont {Gorlach}\ \emph {et~al.}(2018)\citenamefont
  {Gorlach}, \citenamefont {Di~Liberto}, \citenamefont {Recati}, \citenamefont
  {Carusotto}, \citenamefont {Poddubny},\ and\ \citenamefont
  {Menotti}}]{gorlach2018simulation}%
  \BibitemOpen
  \bibfield  {author} {\bibinfo {author} {\bibfnamefont {M.~A.}\ \bibnamefont
  {Gorlach}}, \bibinfo {author} {\bibfnamefont {M.}~\bibnamefont {Di~Liberto}},
  \bibinfo {author} {\bibfnamefont {A.}~\bibnamefont {Recati}}, \bibinfo
  {author} {\bibfnamefont {I.}~\bibnamefont {Carusotto}}, \bibinfo {author}
  {\bibfnamefont {A.~N.}\ \bibnamefont {Poddubny}}, \ and\ \bibinfo {author}
  {\bibfnamefont {C.}~\bibnamefont {Menotti}},\ }\href@noop {} {\bibfield
  {journal} {\bibinfo  {journal} {Phys. Rev. A}\ }\textbf {\bibinfo {volume}
  {98}},\ \bibinfo {pages} {063625} (\bibinfo {year} {2018})}\BibitemShut
  {NoStop}%
\bibitem [{\citenamefont {Johansson}\ \emph {et~al.}(2012)\citenamefont
  {Johansson}, \citenamefont {Nation},\ and\ \citenamefont {Nori}}]{qutip1}%
  \BibitemOpen
  \bibfield  {author} {\bibinfo {author} {\bibfnamefont {J.}~\bibnamefont
  {Johansson}}, \bibinfo {author} {\bibfnamefont {P.}~\bibnamefont {Nation}}, \
  and\ \bibinfo {author} {\bibfnamefont {F.}~\bibnamefont {Nori}},\ }\href
  {\doibase 10.1016/j.cpc.2012.02.021} {\bibfield  {journal} {\bibinfo
  {journal} {Comput. Phys. Commun.}\ }\textbf {\bibinfo {volume} {183}},\
  \bibinfo {pages} {1760} (\bibinfo {year} {2012})}\BibitemShut {NoStop}%
\bibitem [{\citenamefont {Johansson}\ \emph {et~al.}(2013)\citenamefont
  {Johansson}, \citenamefont {Nation},\ and\ \citenamefont {Nori}}]{qutip2}%
  \BibitemOpen
  \bibfield  {author} {\bibinfo {author} {\bibfnamefont {J.}~\bibnamefont
  {Johansson}}, \bibinfo {author} {\bibfnamefont {P.}~\bibnamefont {Nation}}, \
  and\ \bibinfo {author} {\bibfnamefont {F.}~\bibnamefont {Nori}},\ }\href
  {\doibase 10.1016/j.cpc.2012.11.019} {\bibfield  {journal} {\bibinfo
  {journal} {Comput. Phys. Commun.}\ }\textbf {\bibinfo {volume} {184}},\
  \bibinfo {pages} {1234} (\bibinfo {year} {2013})}\BibitemShut {NoStop}%
\bibitem [{\citenamefont {Bohigas}\ \emph {et~al.}(1984)\citenamefont
  {Bohigas}, \citenamefont {Giannoni},\ and\ \citenamefont
  {Schmit}}]{bohigas1984characterization}%
  \BibitemOpen
  \bibfield  {author} {\bibinfo {author} {\bibfnamefont {O.}~\bibnamefont
  {Bohigas}}, \bibinfo {author} {\bibfnamefont {M.-J.}\ \bibnamefont
  {Giannoni}}, \ and\ \bibinfo {author} {\bibfnamefont {C.}~\bibnamefont
  {Schmit}},\ }\href@noop {} {\bibfield  {journal} {\bibinfo  {journal} {Phys.
  Rev. Letters}\ }\textbf {\bibinfo {volume} {52}},\ \bibinfo {pages} {1}
  (\bibinfo {year} {1984})}\BibitemShut {NoStop}%
\bibitem [{\citenamefont {Zimmermann}\ \emph {et~al.}(1986)\citenamefont
  {Zimmermann}, \citenamefont {Meyer}, \citenamefont {K{\"o}ppel},\ and\
  \citenamefont {Cederbaum}}]{zimmermann1986manifestation}%
  \BibitemOpen
  \bibfield  {author} {\bibinfo {author} {\bibfnamefont {T.}~\bibnamefont
  {Zimmermann}}, \bibinfo {author} {\bibfnamefont {H.-D.}\ \bibnamefont
  {Meyer}}, \bibinfo {author} {\bibfnamefont {H.}~\bibnamefont {K{\"o}ppel}}, \
  and\ \bibinfo {author} {\bibfnamefont {L.}~\bibnamefont {Cederbaum}},\
  }\href@noop {} {\bibfield  {journal} {\bibinfo  {journal} {Phys. Rev. A}\
  }\textbf {\bibinfo {volume} {33}},\ \bibinfo {pages} {4334} (\bibinfo {year}
  {1986})}\BibitemShut {NoStop}%
\bibitem [{\citenamefont {Livan}\ \emph {et~al.}(2018)\citenamefont {Livan},
  \citenamefont {Novaes},\ and\ \citenamefont {Vivo}}]{livan2018introduction}%
  \BibitemOpen
  \bibfield  {author} {\bibinfo {author} {\bibfnamefont {G.}~\bibnamefont
  {Livan}}, \bibinfo {author} {\bibfnamefont {M.}~\bibnamefont {Novaes}}, \
  and\ \bibinfo {author} {\bibfnamefont {P.}~\bibnamefont {Vivo}},\ }\href@noop
  {} {\emph {\bibinfo {title} {Introduction to random matrices: theory and
  practice}}},\ Vol.~\bibinfo {volume} {26}\ (\bibinfo  {publisher}
  {Springer},\ \bibinfo {year} {2018})\BibitemShut {NoStop}%
\bibitem [{\citenamefont {Bruder}\ \emph {et~al.}(1993)\citenamefont {Bruder},
  \citenamefont {Fazio},\ and\ \citenamefont
  {Sch{\"o}n}}]{bruder1993superconductor}%
  \BibitemOpen
  \bibfield  {author} {\bibinfo {author} {\bibfnamefont {C.}~\bibnamefont
  {Bruder}}, \bibinfo {author} {\bibfnamefont {R.}~\bibnamefont {Fazio}}, \
  and\ \bibinfo {author} {\bibfnamefont {G.}~\bibnamefont {Sch{\"o}n}},\
  }\href@noop {} {\bibfield  {journal} {\bibinfo  {journal} {Phys. Rev. B}\
  }\textbf {\bibinfo {volume} {47}},\ \bibinfo {pages} {342} (\bibinfo {year}
  {1993})}\BibitemShut {NoStop}%
\bibitem [{\citenamefont {Park}(2012)}]{park2012classical}%
  \BibitemOpen
  \bibfield  {author} {\bibinfo {author} {\bibfnamefont {D.}~\bibnamefont
  {Park}},\ }\href@noop {} {\emph {\bibinfo {title} {Classical dynamics and its
  quantum analogues}}}\ (\bibinfo  {publisher} {Springer Science \& Business
  Media},\ \bibinfo {year} {2012})\BibitemShut {NoStop}%
\end{thebibliography}%

\end{document}

\preprint{AIP/123-QED}
	
	\title[\mytitile]{\mytitile\\~}
	
\maketitle

\section{Linear regime} \label{sec:app_linear}

In the linear regime with no pure dephasing, the Langevin equations for the steady state read:
\begin{equation}
\begin{aligned}
0 &= i(\omega_1 - \omega_d)\hat b_1^\dag - \frac{\gamma_1}{2} \hat b_1^\dag + i J\hat b_2^\dag + \sqrt{\gamma_1}b_{in}^\dag,\\
0 &= i(\omega_2 - \omega_d)\hat b_2^\dag - \frac{\gamma_2}{2} b_{2}^\dag + i J\hat b_{1}^\dag + i J\hat b_{3}^\dag,\\
0 &= i(\omega_3 - \omega_d)\hat b_3^\dag - \frac{\gamma_3}{2} \hat b_{3}^\dag + i J\hat b_{2}^\dag + i J\hat b_{4}^\dag,\\
0 &= i(\omega_4 - \omega_d)\hat b_4^\dag -\frac{\gamma_4}{2} \hat b_{4}^\dag + i J\hat b_{3}^\dag + i J\hat b_{5}^\dag,\\
0 &= i(\omega_5 - \omega_d)\hat b_5^\dag - \frac{\gamma_5}{2} \hat b_5^\dag + i J\hat b_4^\dag,
\end{aligned} 
\end{equation}
omitting zero-photon fluctuations. The solution of this system used for the model curves in Fig. 3~(a) of the main paper is too complex to display; however, for the degenerate case it becomes rather simple:
\begin{equation}
\hat b_{{5}}^\dag={\frac {4\,i{J}^{4}\sqrt {\Gamma} \hat b_{in}^\dag}{ \left( i\delta\,
		\Gamma+2\,{\delta}^{2}-2\,{J}^{2} \right)  \left( i{\delta}^{2}
		\Gamma-2\,i{J}^{2}\Gamma+2\,{\delta}^{3}-6\,\delta\,{J}^{2
		} \right) }}.\label{eq:s21}
\end{equation}
Here, $\delta = \omega_d - \omega$. For the case of strong coupling, $J\gg \Gamma$, one can find that the poles of this expression are at $0, \pm J, \pm \sqrt 3 J $. This is a particular case of a more general statement considering the crystal dispersion relation: for a system of size $N$, the crystal momentum takes $N$ values $k = \frac{2 \pi}{N+1} m,\ m=\pm 1, \pm 2... m \leq N/2,\ m\in \mathbb{Z}\  \cup\ {0}$ if $ N $ is odd. Then the corresponding dispersion relation is $E/\hbar = \omega + 2 J \sin k/2$.

Using the Taylor expansion near each of the poles of \eqref{eq:s21}, one can express is in the Lorentzian form and extract its width in the limit of $J \gg \Gamma$. This procedure will lead to the values $\Gamma/6,\ \Gamma/2,\ 2\Gamma/3$ for the peaks at the detunings $\pm \sqrt{3 J},\ \pm J, 0$, respectively. The sum of the widths equals to $2\Gamma$ and is conserved even when the transmons are arbitrarily detuned from the resonance. We also derive these values in another way in Section III, see below.

\section{Two-qutrit analytical solution}
\begin{figure}
	\includegraphics[width=0.5\linewidth]{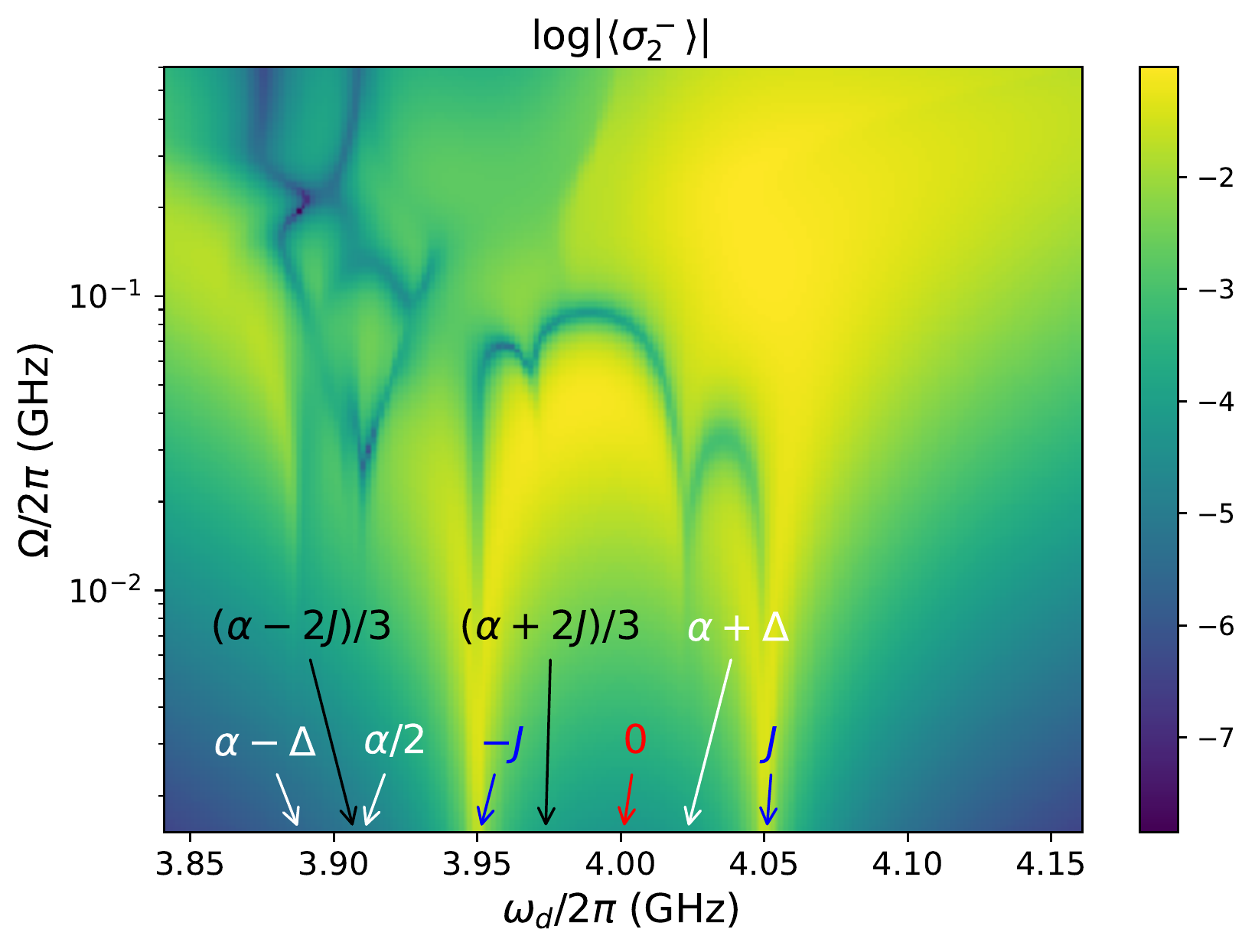}
	\caption{Analytical solution using the expansion of the density matrix. Arrows denote the identified transitions, where $\Delta = \sqrt{16 J^2 + \alpha^2}$, $\omega/2\pi = 4$ GHz, $\alpha/\pi = -181$ MHz, $J/2\pi = 50$ MHz, $\gamma \approx 3.2\, \mu\text{s}^{-1}$.}
	\label{fig:map-sminus-2qb3-log}
\end{figure}

It is instructive to discuss also a simple model of a chain of two identical three-level transmons; this case can be solved analytically. However, even for a pair of qutrits the exact analytical expression for the steady state is too cumbersome work with. Therefore we suggest to analyze its series expansion with respect to the driving amplitude. In the leading order in $\Omega$, the density matrix elements in the steady state can be represented as
$$
\rho_{\text{ss}}^{ij|i'j'} \propto  \Omega^{i+j+i'+j'-4}\, \ket{i,j}\bra{i',j'},
$$
here variables $i$, $j$, $i'$, $j'$ run through the values 1, 2, 3 related to the number of the transmon energy level. We split all elements of the density matrix into groups according to its leading order $n$ in $\Omega$. We start from the case $n=0$. In this group there is only single element related to the transmon's ground state, so for this element we have $\rho^{11|11}_{\text{ss}}=1$. As the next step, $n=1$, there are four linear equations for density matrix elements $\rho^{11|12}_{\text{ss}}$, $\rho^{11|21}_{\text{ss}}$, $\rho^{12|11}_{\text{ss}}$ and $\rho^{21|11}_{\text{ss}}$. Here we take into account the result at the previous step  $\rho^{11|11}_{\text{ss}}=1$. We continue this iterative procedure for $n=2, 3, \dots$ and find the remaining density matrix elements. Thus, we finally find the series expansion for 

$$
\sigma^-_{2}
=
\mathbbm{1} \otimes \begin{pmatrix}
0 & 0 & 0
\\
\sqrt{2}& 0 & 0
\\
0 & 1 & 0
\end{pmatrix}
$$
as
\begin{gather*}
\langle\sigma_{2}^- \rangle
=\,
\text{Tr}\left [ \sigma_{2}  \rho_{\rm ss}\right]
=
\rho^{11|21}_{\rm ss}
+
\rho^{22|21}_{\rm ss}
+
\rho^{33|21}_{\rm ss}
+
\sqrt{2} \left(
\rho^{11|32}_{\rm ss}
+
\rho^{22|21}_{\rm ss}
+
\rho^{33|21}_{\rm ss}
\right)
=\\
-
\frac{16 \Omega J}{16 J^2 + (\Gamma - 4 i \delta)^2}
\\
-\frac{4096 \Omega^3 J^3 (\alpha -i \Gamma -4 \delta )}{(2 \alpha -i \Gamma \
	-4 \delta ) \left(16 J^2+(\Gamma -4 i \delta )^2\right) \left(16 \
	J^2+(\Gamma +4 i \delta )^2\right) \left(16 J^2+(\Gamma -4 i \delta ) \
	(2 i \alpha +\Gamma -4 i \delta )\right)}\\
\vdots
\end{gather*}
where $\delta = \omega_d - \omega$ is the detuning. We stress that in this expansion each element is calculated in its leading order with respect to $\Omega$. For the sake of simplicity, we consider the case $\Gamma \to 0$. In this limit, the term linear in $\Omega$ has two poles $\delta = \pm J$ which we attribute to single-photon resonances. The third order term has three additional poles $\delta = \alpha/2$, $\delta =(\alpha \pm \sqrt{16 J^2 + \alpha^2})/4$ which can be attributed to the two-photon processes. In the fifth order there will be additional three-photon pole at
$\delta = (\alpha \pm 2J)/3$. The result of the numerical calculation of $\langle\sigma_{2}^-\rangle$ for different $\Omega$ and $\omega$ is represented in Fig. \ref{fig:map-sminus-2qb3-log}. As one can see, with the increase of driving power, when driving power become of order of damping, dips appear at frequencies we have determined. This feature is expectable for multi-photonic processes and it is due to effective inverse occupation of the third transmon levels. It occurs when the occupancy of the transmon third level exceeds a second level occupancy, that results in effective suppression of a single-photon process.

There is another approach to find the resonant frequencies responsible for multiphoton processes. 
the Hamiltonian in the rotating frame does not depend on time explicitly. In addition it conserves excitation number, $N$, so it can be split into independent sectors with particular $N$ which we denote as $H_{(N)}$.
In each sector we find eigenenergies:
\begin{widetext}
	$$
	\begin{array}{ccc}
	H_{(1)} = \begin{pmatrix}
	-\delta & J
	\\
	J & -\delta
	\end{pmatrix},
	\qquad &
	\delta = \pm J;
	\\[1em]
	H_{(2)} = \begin{pmatrix}
	\alpha - 2\delta & \sqrt{2} J & 0
	\\
	\sqrt{2} J & - 2\delta & \sqrt{2} J
	\\
	0 & \sqrt{2} J & \alpha - 2 \delta
	\end{pmatrix},
	\qquad &
	\displaystyle
	\delta = \frac{\alpha \pm \sqrt{16 J^2 + \alpha^2}}{4}, \quad \delta = \frac{\alpha}{2};
	\\[2em]
	H_{(3)} = \begin{pmatrix}
	\alpha - 3\delta & 2J
	\\
	2 J & \alpha - 3 \delta
	\end{pmatrix},
	\qquad & \displaystyle
	\delta = \frac{\alpha \pm 2 J}{3};
	\\[2em]
	H_{(4)} = \begin{pmatrix}
	2 \alpha - 4\delta
	\end{pmatrix},
	\qquad & \displaystyle
	\delta = \frac{\alpha}{2}.
	\end{array}
	$$
\end{widetext}\onecolumngrid
This result illustrates why there is no specific poles in series expansion of $\langle\sigma_{2}^-\rangle$ in seventh order: four-photon process is hidden by a two-photon process since both of them have the same energy. This accidental coincidence is a feature of this particular configuration with a couple of three-level systems. For example, if we treat transmons as  four-level systems, then this effect will not occur and four-photon process will be distinguishable.

\section{Weak driving: transmons as two-level systems}

For a single two-level system under the coherent irradiation, electromagnetic dressing in a nonlinear regime gives rise to the splitting of the transmission peak. It is of interest to address transmon chain treating it as a chain of qubits under weak pumping that is expected to be adequate in a weakly nonlinear regime.

We thus consider a XY chain in the transverse field, which is described
by the Hamiltonian
\begin{eqnarray}
H = \omega \sum_{i=1}^{5} \sigma_i^{+} \sigma_i^{-} + J
\sum_{i=1}^{5}(\sigma_i^{+} \sigma_{i+1}^{-} + \sigma_{i+1}^{+}
\sigma_i^{-}),
\label{Hamiltonian}
\end{eqnarray}
where $\omega$ is the excitation energy of each qubit, which corresponds to the transition from the ground to the first excited state of the transmon, and $\sigma_i^{+}$, $\sigma_i^{-}$ are Pauli operators acting in the space of two states of $i$'th qubit.

The Hamiltonian (\ref{Hamiltonian}) conserves excitation number. We readily find eigenstates of this Hamiltonian in the sector with the single excitation. The eigenenergies $E_n$ and eigenstates $S_n^{\dagger} |0\rangle$ are given by
\begin{eqnarray}
E_1 = \omega - J \sqrt{3}, &&
S_1^{\dagger}=\frac{1}{2\sqrt{3}}\sigma_1^{+} - \frac{1}{2}\sigma_2^{+} + \frac{1}{\sqrt{3}}\sigma_3^{+} - \frac{1}{2}\sigma_4^{+} + \frac{1}{2\sqrt{3}}\sigma_5^{+},
\\ \nonumber
E_2 = \omega - J , &&
S_2^{\dagger}=\frac{1}{2}\left(\sigma_1^{+} - \sigma_2^{+} + \sigma_4^{+} - \sigma_5^{+}\right),
\\ \nonumber
E_3 = \omega, &&
S_3^{\dagger}=\frac{1}{\sqrt{3}}\left(\sigma_1^{+} - \sigma_3^{+} + \sigma_5^{+}\right),
\\ \nonumber
E_4 = \omega + J , &&
S_2^{\dagger}=\frac{1}{2}\left(\sigma_1^{+} + \sigma_2^{+} - \sigma_4^{+} - \sigma_5^{+}\right),
\\ \nonumber
E_5 = \omega + J \sqrt{3}, &&
S_5^{\dagger}=\frac{1}{2\sqrt{3}}\sigma_1^{+} + \frac{1}{2}\sigma_2^{+} + \frac{1}{\sqrt{3}}\sigma_3^{+} + \frac{1}{2}\sigma_4^{+} + \frac{1}{2\sqrt{3}}\sigma_5^{+}.
\label{Eigen}
\end{eqnarray}
We assume that energy dissipation rates $\Gamma$ are nonzero only for the rightmost and leftmost qubits of the chain and they are equal to each other. In this case, the energy dissipation rates corresponding to the new state $S_n^{\dagger} |0\rangle$ are given by $\Gamma_n=\Gamma \langle 0 |S_n (\sigma_1^{+} \sigma_1^{-} +  \sigma_5^{+} \sigma_5^{-} )S_n^{\dagger} |0\rangle$, from which we find $\Gamma_1=\Gamma_5 = \Gamma /6$, $\Gamma_2=\Gamma_4= \Gamma/2$, $\Gamma_3= 2\Gamma /3$. All these results are in agreement with our conclusions for the linear regime of the Bose-Hubbard chain. We now express $\sigma_1^{+}$ in terms of the new operators as
\begin{eqnarray}
\sigma_1^{\dagger}=\frac{1}{2\sqrt{3}}S_1^{+} + \frac{1}{2}S_2^{+} + \frac{1}{\sqrt{3}}S_0^{+} + \frac{1}{2}S_4^{+} + \frac{1}{2\sqrt{3}}S_5^{+},
\label{sigma1}
\end{eqnarray}
from which we conclude that in the dressed basis the input
characteristics $\Omega_{n}$ for each state reads as $\Omega_{1}=\Omega_{5}=\frac{1}{2\sqrt{3}}\Omega$, $\Omega_{2}=\Omega_{4}=\frac{1}{2}\Omega$, $\Omega_{3}=\frac{1}{\sqrt{3}}\Omega$. In order to treat the dynamics of the chain under the pumping with frequency $\omega_d$,
we switch to the
dressed basis and consider Maxwell-Bloch equations for each of the new
two-level systems in the rotating frame
\begin{equation}
\frac{d \langle S_n^{z} \rangle}{dt} = -2\Gamma_n \left(1+
\langle S_n^{z} \rangle \right) +  \frac{i \Omega_n \sqrt{\Gamma}}{2}
\left(\langle S_n^{-} \rangle  -
\langle S_n^{+} \rangle \right). \label{sigmaz}
\end{equation}
\begin{equation}
\frac{d \langle S_n^{-} \rangle}{dt} = -\left(i (E_n-\omega_d) +
\Gamma_n\right) \langle S_n^{-} \rangle + i
\sqrt{\Gamma} \langle S_n^{z} \rangle \Omega_n.
\label{sigmaminus}
\end{equation}

The steady state solution for $\langle S_n^{-} \rangle$ is
\begin{equation}
\langle S_n^{-} \rangle  = -
\frac{i \Omega_{n} \sqrt{\Gamma}}{i\left(E_n - \omega\right) + \Gamma_n}
\frac{1}{1+ \frac{|\Omega_{n}|^2 \Gamma/2}{\left(E_n - \omega_d\right)^2 +
		\Gamma_n^2}}. \label{sigmaminusst}
\end{equation}
We also express $\sigma_{5}^{+} $ in terms of dressed states operators as \begin{eqnarray}
\sigma_5^{+}=\frac{1}{2\sqrt{3}}S_1^{+} - \frac{1}{2}S_2^{+} + \frac{1}{\sqrt{3}}S_0^{+} - \frac{1}{2}S_4^{+} + \frac{1}{2\sqrt{3}}S_5^{+}.
\label{sigma5}
\end{eqnarray}
Taking into account Eq. (\ref{sigmaminusst}), we obtain $S_{21}$ as a sum of contributions of five states
\begin{eqnarray}
S_{21} = \frac{1}{6} \frac{\Gamma}{i\left(\Delta
	\omega - J \sqrt{3}\right) + \Gamma/6}
\frac{1}{1+ \frac{1}{2}\frac{ \Omega^2 \Gamma /6}{\left(\Delta
		\omega - J \sqrt{3}\right)^2 + \Gamma^2/36}}  \nonumber
-\frac{1}{2} \frac{\Gamma}{i\left(\Delta
	\omega - J \right) + \Gamma/2}
\frac{1}{1+ \frac{1}{2} \frac{ \Omega^2 \Gamma /2}{\left(\Delta
		\omega - J \right)^2 + \Gamma^2/4}} \\ \nonumber
+ \frac{2}{3} \frac{\Gamma}{i\left(\Delta
	\omega \right) + 2\Gamma/3}
\frac{1}{1+ \frac{1}{2}\frac{ 2 \Omega^2 \Gamma /3}{\left(\Delta
		\omega \right)^2 + 4 \Gamma^2/9}} %\\ \nonumber
-\frac{1}{2} \frac{\Gamma}{i\left(\Delta
	\omega + J \right) + \Gamma/2}
\frac{1}{1+ \frac{1}{2}\frac{ \Omega^2 \Gamma /2}{\left(\Delta
		\omega + J \right)^2 + \Gamma^2/4}} \\ %\nonumber
+ \frac{1}{6} \frac{\Gamma}{i\left(\Delta
	\omega + J \sqrt{3}\right) + \Gamma/6}
\frac{1}{1+ \frac{1}{2}\frac{ \Omega^2 \Gamma /6}{\left(\Delta
		\omega + J \sqrt{3}\right)^2 + \Gamma^2/36}},
\label{bout}
\end{eqnarray}
where $\Delta \omega = \omega - \omega_d$.

At very weak drivings, i.e., in linear regime, peaks of $|S_{21}|$ are at eigenenergies of $H$. At stronger drivings,
$ \frac{|\Omega_{n}|^2 \Gamma / 2}{\left(E_n - \omega_d\right)^2 +
	\Gamma_n^2}$ dominates over $1$ and
all denominators of the form $i\left(E_n - \omega_d\right) + \Gamma_n$
disappear from the expression for $S_{21}$, so that
transmission peaks become suppressed and are gradually transformed into local minima. Along the crossover between the two regimes,
individual maxima split into peaks, which are due to the fact that the dressed states of the chain become also dressed with the incident light.

We stress that our simple model is applicable only at low drivings, when at maximum a single excitation is created in the system. Particularly, this model totally misses multiphoton excitation processes, which appear at the same order in $\Omega$ as the nonlinearities we here consider. However, there exists a domain of parameters at weak pumping, where multiphoton processes are not so pronounced, while transmission peaks splitting is discernable, see, e.g., Fig. 2. Particularly, this applies to blue-detuned regimes near resonances as transmons anharmonicities are negative.

\begin{figure}
	\includegraphics[width=.5\linewidth]{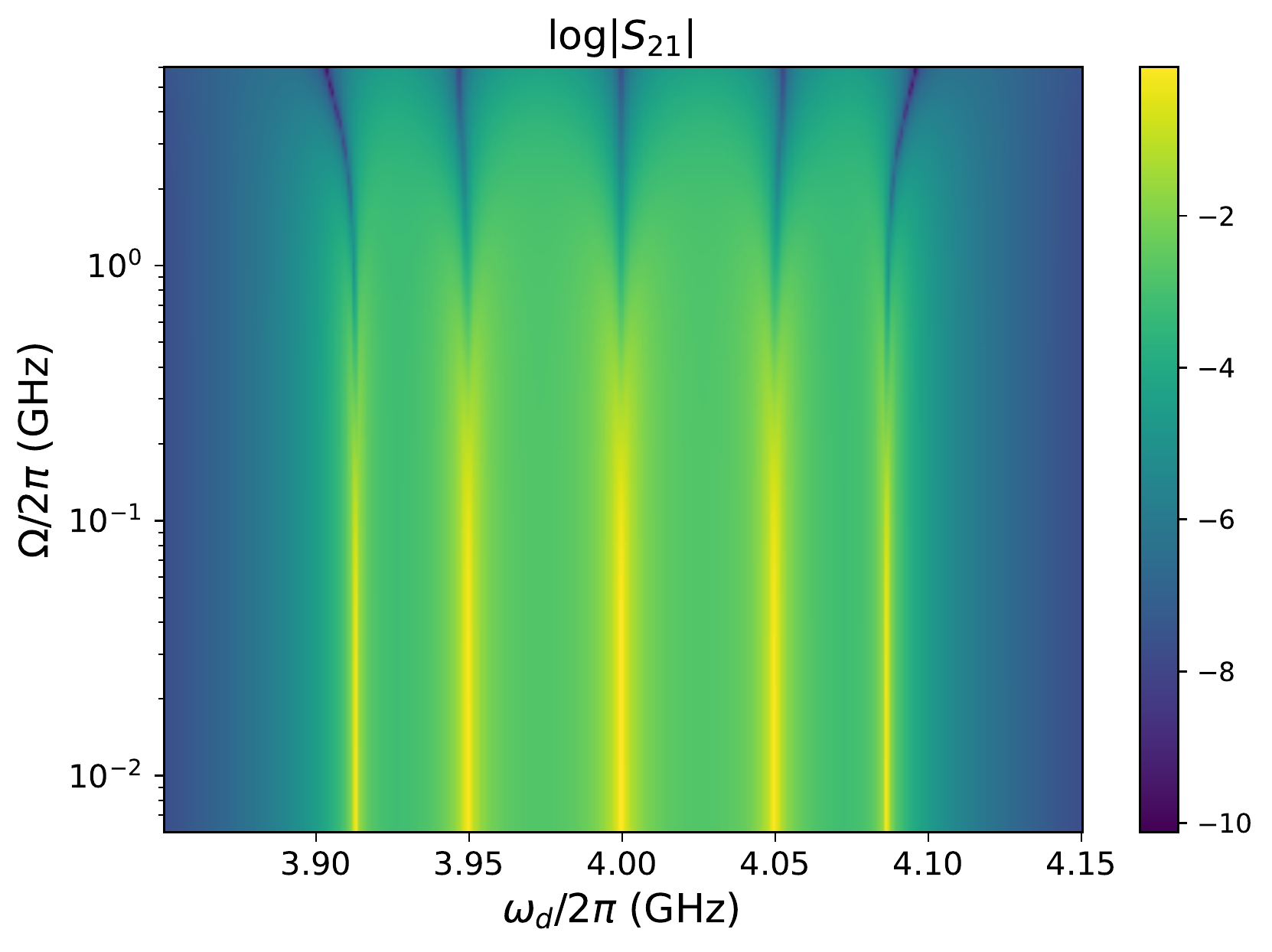}
	\caption{The behavior of $\ln |S_{21}|$ for the chain of qubits as
		a function of $\omega_d$ and $\Omega$.}
	\label{fig:qubitchain}
\end{figure}

\section{Sample fabrication}

To fabricate a five-qubit array we use the process similar to ones described in Ref. [26] of the main text. The process includes two stages: I) double-angle evaporation of a base layer and Josephson junctions (including transmon capacitor ground plane, waveguides, resonators and flux bias lines) followed by lift-off; II) resist-based low impedance crossover fabrication. 

The fabrication process starts with multi-step wet chemical cleaning of a high-resistivity intrinsic silicon sample ($\rho$ > 10000 $\Omega\cdot$cm, 525 $\mu$m thick) in a Piranha solution (1:4) followed by native oxide removal in HF (1:50) for 120 seconds. Than a two-layer e-beam resists stack (100 nm thick AR-P e-beam resist on top of a 700 nm thick MMA copolymer) is spin coated followed by 50kV e-beam exposure. After development and oxygen plasma treatment, we performed e-beam shadow evaporation of Al-AlOx-Al Josephson junctions ($\pm$10.2$^\circ$, 25/45 nm) with Plassys PVD tool. Low impedance free-standing crossovers are fabricated by means of a four-stage [Chen, Z. et al. Fabrication and characterization of aluminum airbridges for superconducting microwave circuits. Appl. Phys. Lett. 104, 052602 (2014)] process: (I) air-bridges pads laser lithography, (II) 300 nm thick Al film e-beam deposition, (III) air-bridges topology laser lithography, (IV) Cl2-based dry plasma etching. Finally, we stripped both resist layers in an NMP-based solvent.

\begin{figure}
	\includegraphics[width=.49\linewidth]{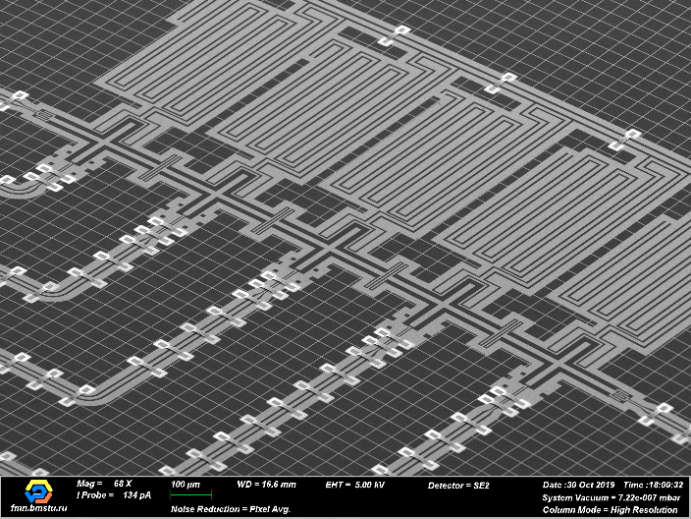}
		\includegraphics[width=.49\linewidth]{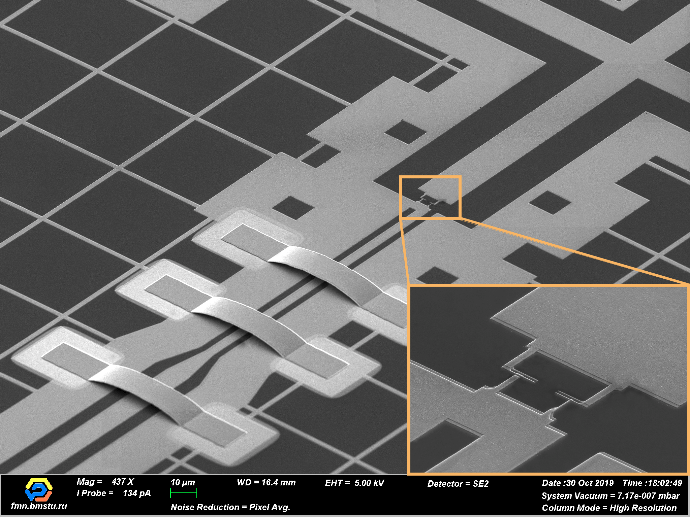}
		\caption{SEM image of a five qubits array (left) and SEM image of a transmon capacitor and zoomed in Josephson junction SQUID (right). }
\end{figure}